\documentstyle[epsf,aps]{revtex}
\renewcommand{\thefootnote}{\fnsymbol{footnote}}

\begin{document}
\newcommand{\be}{\begin{eqnarray}}
\newcommand{\dlq}{\lq\lq}
\newcommand{\ee}{\end{eqnarray}}
\newcommand{\ben}{\begin{eqnarray*}}
\newcommand{\een}{\end{eqnarray*}}
\newcommand{\stackeven}[2]{{{}_{\displaystyle{#1}}\atop\displaystyle{#2}}}
\newcommand{\lsim}{\stackeven{<}{\sim}}
\newcommand{\gsim}{\stackeven{>}{\sim}}
\newcommand{\bas}{\overline{\alpha}_s}
\newcommand{\un}[1]{\underline{#1}}
\renewcommand{\baselinestretch}{1.0}
\newcommand{\as}{\alpha_s}

\newcommand{\sll}{\raise.15ex\hbox{$/$}\kern-.43em\hbox{$l$}}
\newcommand{\slepsilon}{\raise.15ex\hbox{$/$}\kern-.53em\hbox{$\epsilon$}}
\newcommand{\slp}{\raise.1ex\hbox{$/$}\kern-.63em\hbox{$p$}}
\newcommand{\slq}{\raise.1ex\hbox{$/$}\kern-.53em\hbox{$q$}}
\newcommand{\slv}{\raise.1ex\hbox{$/$}\kern-.63em\hbox{$v$}}
\newcommand{\slk}{\raise.15ex\hbox{$/$}\kern-.53em\hbox{$k$}}
\newcommand{\slpartial}{\raise.15ex\hbox{$/$}\kern-.53em\hbox{$\partial$}}

\def\eq#1{{Eq.~(\ref{#1})}}
\def\fig#1{{Fig.~\ref{#1}}}
\begin{flushright}
NT@UW--04--013 \\
INT--PUB--04--15
\end{flushright}
\vspace*{1cm} 
\setcounter{footnote}{1}
\begin{center}
{\Large\bf Inclusive Two--Gluon and Valence
Quark--Gluon~\\[5mm] Production in DIS and pA}
\\[1cm]
Jamal Jalilian-Marian $^{1}$ and Yuri V.\ Kovchegov  $^{2}$ \\ ~~ \\ 
{\it $^1$ Institute for Nuclear
Theory, University of Washington, Box 351550 } \\ {\it Seattle, WA
98195 } \\ ~~ \\
{\it $^{2}$ Department of Physics, University of Washington, 
Box 351560} \\ {\it Seattle, WA 98195 } \\ ~~ \\ ~~ \\
\end{center}
\begin{abstract}
\noindent We calculate production cross sections of a forward 
quark--gluon pair and of two gluons at mid-rapidity in Deep Inelastic
Scattering and in high energy proton--nucleus collisions. The
calculation is performed in the framework of the Color Glass
Condensate formalism.  We first calculate the cross sections in the
quasi-classical approximation, which includes multiple rescatterings
in the target.  We then proceed to include the effects of non-linear
small-$x$ evolution in the production cross sections. It is
interesting to note that our result for the two-gluon production cross
section appears to be in direct violation of AGK cutting rules, which
is the first example of such violation in QCD. The calculated
quark--gluon and gluon--gluon production cross sections can be used to
construct theoretical predictions for two-particle azimuthal
correlations at RHIC and LHC ($I^{p(d)A}$) as well as for Deep
Inelastic Scattering experiments at HERA and eRHIC.
\end{abstract}
\renewcommand{\thefootnote}{\arabic{footnote}}
\setcounter{footnote}{0}

\section{Introduction}

Recent prediction of high-$p_T$ suppression in the nuclear
modification factor $R^{dA}$ at forward rapidity RHIC $dAu$
collisions \cite{klm,kkt1,aaksw} based on the physics of parton
saturation/Color Glass Condensate
\cite{glr,mq,bm,mv,kjkmw,bk,jimwlk,GM} has been confirmed by the
experimental data in \cite{brahms-1,brahms-2,phenix,phobos,star}. The
prediction of \cite{kkt1} was based on the calculation of inclusive
gluon production cross section in DIS and $pA$ collisions. The
calculation was first done in the quasi-classical framework of
McLerran-Venugopalan model including all multiple rescatterings
\cite{km} (see also \cite{kst,dmc,yuri1,kw}). The effects of 
non-linear small-$x$ evolution \cite{bk,jimwlk,Braun} were included in
the obtained formula in \cite{kt} (see also \cite{braun}). In
\cite{kkt1,aaksw} it was argued that at lower energies/rapidities,
where the particle production is given by the quasi-classical formula
from \cite{km}, the nuclear modification factor $R^{pA}$ should
exhibit low-$p_T$ suppression together with a strong enhancement at
high-$p_T$, known as Cronin effect \cite{cronin} (see also
\cite{knst,ag,bkw,jmnv} for similar conclusions). However, at higher
energies/rapidities, when quantum evolution becomes important, one
should expect suppression of $R^{pA}$ at all $p_T$
\cite{klm,kkt1,aaksw} due to the onset of BFKL anomalous dimension for
gluon distributions \cite{bfkl}. (It had been earlier suggested in
\cite{adjjm} that the forward rapidity region would be most sensitive
to small-$x$ evolution effects.) Similar argument about enhancement
and suppression can be carried through for valence quark production
cross section calculated in \cite{adjjm}. The suppression has been
confirmed experimentally in \cite{brahms-1,brahms-2}. The centrality
dependence of the observed suppression was also in agreement with the
predictions of the Color Glass Condensate formalism
\cite{kkt1}. Further developments in the area included an analysis of
running coupling corrections \cite{iit} and a study of similar
suppression in di-lepton production \cite{jjm1} (see also
\cite{boris}). Recently, a more quantitative analyses  \cite{jjm2,kkt2} 
based on the Color Glass Condensate formalism have been performed
which show good agreement with the data \cite{brahms-1,brahms-2}.

Another distinctive prediction of the Color Glass Condensate (CGC) 
\cite{glr,mq,bm,mv,kjkmw,bk,jimwlk} is the disappearance of 
back-to-back jets in the low $p_T < Q_s$ and intermediate $p_T \gsim
Q_s$ transverse momentum regions. While the single particle spectra in
$dAu$ collisions at RHIC have been successfully described by
CGC-inspired models \cite{jjm2,kkt2}, it is important to go beyond
single particle spectra and probe other observables, such as two
particle correlations, in order to map out the region of phase space
where CGC is the dominant physics. The inclusive two particle (gluon)
cross section at high energy is given by the $k_T$-factorization
\cite{lr} in the high $p_T$ region ($p_T \gg Q_s$) with the gluon distribution
function evolving via the BFKL evolution equation
\cite{bfkl}. Models based on $k_T$ factorization have been applied to
many different processes, such as the non-flow contribution to the
elliptic flow observable $v_2$ in heavy ion collisions \cite{ktflow}.
Recently a similar model of two-particle correlations in $dA$ was used
in \cite{klm2} to predict broadening and disappearance of back-to-back
correlations in $pA$ (or $dA$) collisions. The predictions of
\cite{klm2} appear to be confirmed by the preliminary data reported in
\cite{star2}, thus strengthening the case for saturation/Color Glass
Condensate in $dAu$ data at RHIC.

Nevertheless, a theoretically rigorous treatment of inclusive
two-particle production in DIS and proton-nucleus collisions in the
low $p_T$ region ($p_T \lsim Q_s$) has not been performed yet. It is
clearly needed in order to provide reliable predictions in the $p_T
\lsim Q_s$ momentum region, which is the region where the new physics of
CGC is expected to be most pronounced. (Very recently, there has been
a series of articles investigating quark--antiquark production in $pA$
collisions using the quasi-classical approximation in the CGC
formalism \cite{bgv}.)  

Our goal in this work is to derive inclusive two-particle production
cross sections using the Color Glass Condensate formalism. We start by
considering production of two gluons. We assume that the two gluons
are separated by a large rapidity interval so that their respective
rapidities are ordered, $y_2 \gg y_1$. This kinematics is, for
instance, relevant to the case of two particle production in $p(d)A$
collisions (for example at RHIC or LHC) when one of the produced
particles is in the mid-rapidity region while the second particle is
closer to the forward rapidity region. In Sect. II, we derive an
expression for two-gluon inclusive cross section in Deep Inelastic
Scattering (DIS), using the quasi-classical approximation (the
McLerran-Venugopalan model) in the Color Glass Condensate formalism
and making the large-$N_c$ approximation to simplify the calculations.
(The quasi-classical approximation employed here is identical to the
one used in \cite{km,kst,dmc} to describe single gluon production.) 
The result for two-gluon production cross section in the
quasi-classical approximation is given by Eqs. (\ref{disincl}) and
(\ref{2Gcl}). We note that a similar expression for two-quark
production cross section was obtained previously for DIS in
\cite{nszz}.  In Sect. III, we include the effects of nonlinear
small-$x$ evolution \cite{bk} in the two-gluon inclusive cross section
obtained in Sect. II. The final answer for the two-gluon inclusive
cross section for DIS is given in \eq{2Gincl}. This result can be
easily generalized to $pA$ collisions.

An ansatz for two gluon inclusive cross section including saturation
effects was written in \cite{braun} inspired by $k_T$-factorization
together with AGK cutting rules \cite{agk}. We note that the
diagrammatic structure of our answer in \eq{2Gincl} does not seem to
adhere to AGK cutting rules' expectation for two-gluon inclusive cross
section. Furthermore, we are unable to cast the expression
(\ref{2Gincl}) into $k_T$-factorized form used in
\cite{braun}. However, the leading twist $k_T$-factorization
expression \cite{lr,kt} can be reproduced exactly from \eq{2Gincl}, as
will be discussed at the end of Sect. III.

In Sect. IV we calculate the inclusive production of a valence quark
and a gluon in $pA$ collisions both in the quasi-classical
approximation and including the quantum evolution in the target. The
rapidities of the valence quark and the gluon are assumed to be
comparable and large (both quark and gluon are produced in the forward
rapidity region). The result is given by Eqs. (\ref{eq:m_12}- \ref{eq:m_34}).  
These expressions together
with \eq{2Gincl} can be used to describe the nuclear modification
factor for azimuthal correlations $I^{dAu}$ at any rapidity between
mid-rapidity and the deuteron beam at RHIC. In particular,
Eq. (\ref{2Gincl}) provides the theoretical basis for the correlation
analysis carried out in \cite{klm2}.


\section{Two-Gluon Production in the Quasi-Classical Approximation}

\begin{figure}
\begin{center}
\epsfxsize=10cm
\leavevmode
\hbox{\epsffile{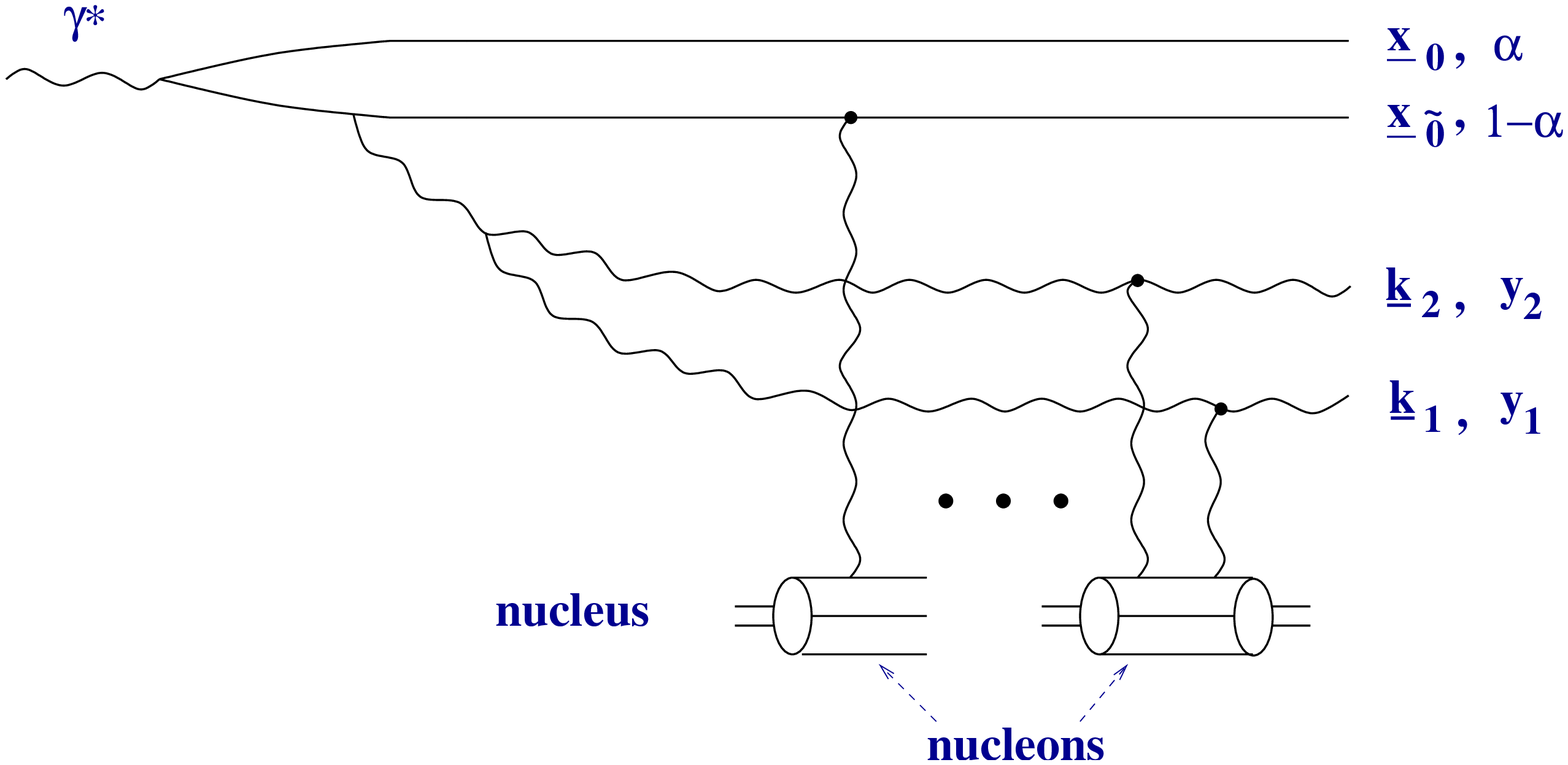}}
\end{center}
\caption{Two-gluon production in DIS on a nucleus including multiple 
rescatterings. }
\label{lodis}
\end{figure}

In this Section we are going to derive an expression for inclusive
two-gluon production cross section in DIS including all multiple
rescatterings of the two produced gluons and the quark-antiquark pair
on the nucleons in the target nucleus \cite{Mueller1,mv,kjkmw}. A
typical diagram contributing to the process is shown in
\fig{lodis}. The two produced gluons have transverse momenta ${\un k}_1$ 
and ${\un k}_2$ and rapidities $y_1$ and $y_2$ correspondingly. To
simplify the calculations we will consider the case when $y_2 \gg
y_1$. A more general case of $y_2 \sim y_1$ was considered in
\cite{LO} for two-gluon production at the leading twist level given by 
$k_T$-factorization. Our goal here is to include the saturation
effects in the two-gluon production cross section, which means summing
all twists. We will achieve this difficult task only for a simpler
case of $y_2 \gg y_1$, though, in principle, the more general case
$y_2 \sim y_1$ presents no new conceptual difficulties and is only
technically more complicated. 

As shown in \fig{lodis} the gluon production process in DIS in the
quasi-classical approximation \cite{Mueller1,mv,kjkmw} consists of two
factorisable stages. First the incoming virtual photon splits into a
quark-antiquark pair, which emits two gluons in the incoming wave
function. (The time scale for this splitting and gluon emissions is
much longer than the time of interaction with the target.) The whole
system multiply rescatters on the nuclear target. (In general the
gluon emissions can happen after the interaction with the target, as
will be discussed shortly.) In the quasi-classical approximation
considered here the interactions with the nucleons are limited to no
more than two exchanged gluons per nucleon (see the second reference
in \cite{kjkmw}). Single gluon production in the same approximation
was calculated for $pA$ collisions in \cite{km} and for DIS in
\cite{yuri1}.

Let us assume that the virtual photon has a large ``$+$'' component of
the momentum and the nucleus had a large ``$-$'' component of its
momentum. Than the diagram of the process shown in \fig{lodis} is
dominant in $A^+ = 0$ gauge. In this Section we will perform all the
calculations in the framework of the light cone perturbation theory in
$A^+ = 0$ gauge \cite{BL}. First of all let us explicitly factor out
the wave function $\Phi^{\gamma^* \rightarrow q\bar q} ({\underline
x}_{0{\tilde 0}}, \alpha)$ of the virtual photon splitting in a
quark-antiquark pair of transverse size ${\un x}_{0{\tilde 0}} = {\un
x}_0 - {\un x}_{\tilde 0}$ with the quark carrying a fraction $\alpha$
of the virtual photon's light cone momentum. The wave function
$\Phi^{\gamma^* \rightarrow q\bar q} ({\underline x}_{0{\tilde 0}},
\alpha)$ is a well-known function and can be found, for example, in
\cite{NZ,KMc}. The two-gluon inclusive production cross section can be
written as
\be\label{disincl}
\frac{d \sigma^{\gamma^* A \rightarrow q{\bar q}GGX}}{d^2 k_1 \, 
dy_1 \, d^2 k_2 \, dy_2} \, = \, \frac{1}{2 \pi^2} \, \int \, d^2
x_{0{\tilde 0}} \, \int_0^1 d \alpha \,
\Phi^{\gamma^* \rightarrow q\bar q} ({\underline x}_{0{\tilde 0}}, \alpha) \, 
\frac{d {\hat \sigma}^{q{\bar q}A\rightarrow q{\bar q}GGX}}{d^2 k_1 \, 
dy_1 \, d^2 k_2 \, dy_2}({\underline x}_{0{\tilde 0}}).
\ee


\subsection{Time-Ordering Rules}

To calculate two-gluon production cross section for a quarkonium
scattering on a nucleus, similarly to \cite{km,yuri1} one has to
consider various possible ordering of the emissions of the two gluons
by the $q\bar q$ pair. The interaction with the nucleus target can be
considered instantaneous compared to long lifetimes of emitted
gluons. Thus we will denote the moment of interaction with the target
by the light cone time $\tau \equiv x^+ = 0$. If $\tau_1$ and $\tau_2$
are the times of the emission of the two gluons, the possible emission
ordering in the amplitude reduces to three cases: (i) both gluons are
emitted before the interaction, $\tau_1 , \tau_2 <0$; (ii) one gluon
is emitted before the interaction and the other one is emitted after
the interaction, $\tau_1 < 0 < \tau_2$ or $\tau_2 < 0 < \tau_1$; (iii)
both gluons are emitted after the interaction, $\tau_1, \tau_2
>0$. 

\begin{figure}
\begin{center}
\epsfxsize=10cm
\leavevmode
\hbox{\epsffile{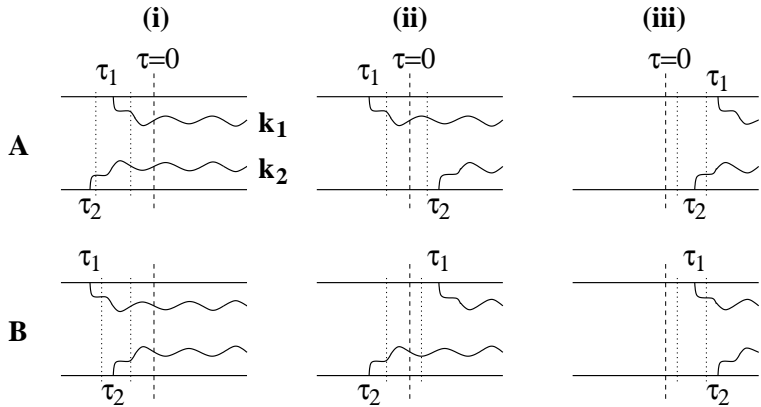}}
\end{center}
\caption{Possible orderings of the emission of the two gluons by the 
quark-antiquark pair.}
\label{order}
\end{figure}

The three cases are represented in \fig{order} for a particular
coupling of the two gluons to the $q\bar q$ pair. There the dashed
line in the middle denotes the (instantaneous) interaction with the
target. The dashed line comprises {\sl all} the multiple rescatterings
like the ones shown in \fig{lodis}. The dotted lines represent
intermediate states, which will give energy denominators in light cone
perturbation theory \cite{BL}. Even though \fig{order} shows a
particular way of the gluons' coupling to the $q\bar q$ pair, the
conclusions we will draw below about which diagrams dominate will be
applicable to other couplings of the gluons.

Let us define the light cone energy of a gluon or a quark line
carrying momentum (${\un k}$, $k^+$) as \cite{BL}
\be
E_k \, \equiv \, k^- \, = \, \frac{{\un k}^2}{2 \, k^+}. 
\ee
In Regge kinematics that we consider here the light cone momenta of
the gluons are ordered, such that $k_2^+ \gg k_1^+$ and $E_{k_2} \ll
E_{k_1}$. First we consider case (i) in \fig{order}. The diagrams (i)A
and (i)B (top and bottom) are different only by energy
denominators. Therefore, forgetting the rest of the diagram for now,
we write
\be\label{iA}
(i)A \, \sim \, \frac{1}{E_{k_2}} \, \frac{1}{E_{k_1} + E_{k_2}} \,
\approx \, \frac{1}{E_{k_2}} \, \frac{1}{E_{k_1}}
\ee
and
\be\label{iB}
(i)B \, \sim \, \frac{1}{E_{k_1}} \, \frac{1}{E_{k_1} + E_{k_2}} \,
\approx \, \frac{1}{E_{k_1}^2}.
\ee
The intermediate states giving the energy denominators in
Eqs. (\ref{iA}) and (\ref{iB}) are shown by dotted lines in
\fig{order}.  Since $E_{k_2} \ll E_{k_1}$, Eqs. (\ref{iA}) and
(\ref{iB}) imply that $(i)A \gg (i)B$. Therefore the diagram (i)B can
be safely neglected. The conclusion we draw from this analysis is that
for gluon emissions before the interaction ($\tau_1, \tau_2 < 0$) the
(longitudinally) harder gluon has to be emitted first, as pictured in
the diagram (i)A in \fig{order}.

In calculating the diagrams (i)A and (i)B we have neglected the
``$-$'' component of the momenta of the $q\bar q$ pair, because they
are negligibly small. The quark and the antiquark carry a very large
``$+$'' component of the momentum, of the order of $p^+ \gg k_2^+ \gg
k_1^+$, which leads to negligibly small light cone energy $E_p$. We
have also neglected the change in the ``$-$'' component of the target
momentum, since the interaction with the target took place after the
intermediate states which gave the energy denominators in case (i) in
\fig{order}. This is not the case in the rest of the diagrams in 
\fig{order}. The ``$-$'' momentum/light cone energy of the target 
changes due to the interaction (dashed line). However, since the light
cone energy is conserved in the final state ($\tau = + \infty$), the
change of the target's ``$-$'' momentum is compensated by the change
of the ``$-$'' momentum of the projectile, which is mostly due to
appearance of two extra gluons leading to an addition of extra
$E_{k_1} + E_{k_2}$ to the $q\bar q$ wave function's light cone
energy. Therefore, the target's light cone energy decreases by
$E_{k_1} + E_{k_2}$ after the interaction. Thus, when calculating the
energy denominators of the intermediate states after the interaction
one has to add the change in the light cone energy of the target to
the energies of the lines shown in \fig{order}. This is equivalent to
subtracting $E_{k_1} + E_{k_2}$ in the corresponding energy
denominators. (This rule is worked out in more detail in Sect. IIIA of
\cite{kt}.)

Guided by the rule we just derived we write for the energy
denominators of the diagrams in the case (ii) in \fig{order}
\be\label{iiA}
(ii)A \, \sim \, \frac{1}{E_{k_1}} \, \frac{1}{E_{k_1} - (E_{k_1} +
E_{k_2})} \, = \, - \frac{1}{E_{k_1}} \, \frac{1}{E_{k_2}},
\ee
\be\label{iiB}
(ii)B \, \sim \, \frac{1}{E_{k_2}} \, \frac{1}{E_{k_2} - (E_{k_1} +
E_{k_2})} \, = \, - \frac{1}{E_{k_2}} \, \frac{1}{E_{k_1}}.
\ee
As we see from Eqs. (\ref{iiA}) and (\ref{iiB}) the two diagrams are
of the same order, $(ii)A \sim (ii)B \ [\sim (i)A]$, and neither of
them can be neglected. (As we will see below diagrams $(ii)A$ and
$(ii)B$ are different in the parts responsible for the interaction
with the target, so while being parametrically of the same order, they
are not identically equal.)

Finally, calculating the graphs in the case (iii) of \fig{order} one
arrives at
\be\label{iiiA}
(iii)A \, \sim \, \frac{1}{E_{k_1} + E_{k_2}} \, \frac{1}{E_{k_1}} \,
\approx \, \frac{1}{E_{k_1}^2},
\ee
and
\be\label{iiiB}
(iii)B \, \sim \, \frac{1}{E_{k_1} + E_{k_2}} \, \frac{1}{E_{k_2}} \,
\approx \, \frac{1}{E_{k_1}} \, \frac{1}{E_{k_2}}.
\ee
Since $E_{k_1} \gg E_{k_2}$ we conclude that $(iii)A \ll (iii)B$. The
diagram (iii)A should be neglected. Therefore, we derive a rule for
late-time emissions, which take place after the interaction ($\tau_1,
\tau_2 >0$): the harder gluon has to be emitted {\sl after} the softer 
gluon. It is interesting to note that this ordering is the exact
inverse of the ordering giving the leading contribution at early times
before the interaction. The rule can also be generalized to any number
of gluon emissions contributing to the BFKL \cite{bfkl} or,
equivalently, dipole evolution \cite{dip,bk}: in the evolution at
early times preceding the interaction the gluons are ordered so that
the harder gluons are emitted before the softer ones \cite{dip}. The
ordering is reversed for late times following the interaction, where
the harder gluons should be emitted after the softer gluons to pick up
the leading logarithmic contribution.  This observation was made
previously in \cite{kt}.


\subsection{Two-Gluon Inclusive Cross Section in the Quasi-Classical 
Approximation}

\begin{figure}
\begin{center}
\epsfxsize=15cm
\leavevmode
\hbox{\epsffile{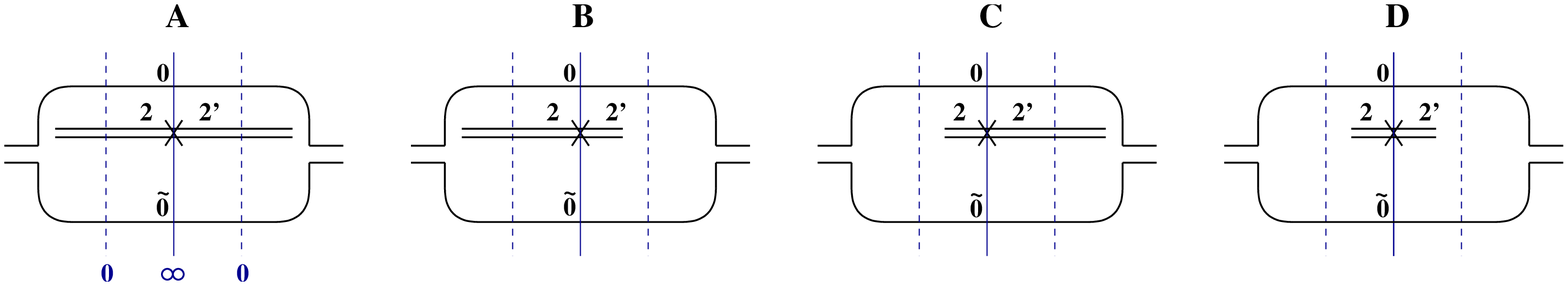}}
\end{center}
\caption{All possible emission of the harder gluon $\# 2$ by the 
$q\bar q$ pair.}
\label{em2}
\end{figure}

The diagrams contributing to emission of the harder gluon with
momentum (${\un k}_2, y_2$) are shown in \fig{em2}. (In the following
we will refer to this gluon as gluon $\# 2$ and to the other (softer)
produced gluon as gluon $\# 1$.) To simplify the color algebra we will
continue the calculation in 't Hooft's large-$N_c$ limit. Only planar
diagrams will contribute for gluon emission. Using the notation from
Mueller's dipole model \cite{dip,CM} we denote the gluon in the
large-$N_c$ limit by a double quark line and leave the ends of the
gluon line disconnected from the quark lines. The latter notation
indicates a sum over all possible connections of the gluon to the
$q\bar q$ pair.

As in \fig{order}, the dashed lines in \fig{em2} denote the $\tau = 0$
moment of the interaction of the system with the target
nucleus. However, unlike \fig{order}, in \fig{em2} we depict the
squares of the amplitude contributing to the total production cross
section. Therefore each diagram has two dashed lines corresponding to
interaction with the target in the amplitude and in the complex
conjugate amplitude. The solid vertical lines denote the final state
at $\tau = + \infty$.

Similar to Mueller's dipole model \cite{dip} the emitted gluon $\# 2$
in \fig{em2}A splits the dipole $0{\tilde 0}$ into two color
dipoles. The emission of the softer gluon $\# 1$ can happen in any of
these two dipoles. However, the original dipole model \cite{dip} was
written for the calculation of the total cross sections, where one
only has to calculate the forward scattering amplitude of the
quarkonium. In that quantity all the final state emissions ($\tau >
0$) cancel, as was shown in \cite{CM}. This is not the case for the
inclusive production cross section that we want to calculate here. All
final state emissions has to be taken into account, as shown in
\fig{em2}. Also, since we are interested in gluon production, the
momentum of gluon $\# 2$ is fixed. Therefore, since we are going to
perform our calculations in transverse coordinate space, we have to
keep the transverse coordinates of the gluon $\# 2$ different on both
sides of the cut. (To obtain the cross section we will afterwards
perform a Fourier transform into transverse momentum space.) Thus, the
gluon's transverse coordinate is denoted by ${\un x}_2$ to the left of
the cut and ${\un x}_{2'}$ to the right of the cut. Then the color
``dipole'' formed by, say, the lines $2$, $2'$ and ${\tilde 0}$ in
\fig{em2}A would not be literally a dipole since one needs more than
two transverse coordinates to describe it, but it would still have the
color topology of a dipole and we will refer to it as a ``dipole''
below.

Let us start by analyzing the gluon production in \fig{em2}A. As was
mentioned before, the softer gluon $\# 1$ can be emitted either off
the color ``dipole'' formed by lines $0$, $2$ and $2'$ or off the
``dipole'' $2,2',{\tilde 0}$. In the following analysis we will
concentrate on the latter case of emission of gluon $\# 1$ in
``dipole'' $2,2',{\tilde 0}$. (A generalization to emission in
``dipole'' $0$, $2$ and $2'$ is straightforward.) We will denote by
$M_0 ({\un x}_2, {\un x}_{2'}, {\un x}_{\tilde 0}; {\un k}_1)$ the
cross section of emission of a softer gluon $\# 1$ in the ``dipole''
$2,2',{\tilde 0}$. Than the ``dipole'' $0,2,2'$ would not have gluon
emissions in it, but it would still be able to interact with the
target. Interactions of the target with the line $0$ would cancel due
to real-virtual cancellations \cite{km,KMc,kt}. Interactions with the
lines $2$ and $2'$ do not cancel: instead they are given by the
S-matrix of a $22'$ quark dipole interacting with the target
\cite{kt,BDMS}. The S-matrix is given by 
\be\label{S0}
S_0 ({\un x}_{2}, {\un x}_{2'}) = 1 - N_0 ({\un x}_{2}, {\un x}_{2'}), 
\ee
where, in the Glauber-Mueller approximation, the forward scattering
amplitude $N_0$ is \cite{Mueller1}
\be
\label{N0}
N_0 ({\un x}_{2}, {\un x}_{2'}) \, = \, 1 - e^{- {x}_{22'}^2
Q_{s0}^2 \, \ln (1/{x}_{22'} \Lambda )/4},
\ee
where ${x}_{22'} = |{\un x}_{2} - {\un x}_{2'}|$ and the quark
saturation scale in the McLerran-Venugopalan model $Q_{s0}$
\cite{mv,kjkmw} is given by (in the large-$N_c$ limit)
\be\label{sat}
Q_{s0}^2({\un b})\,=\, 2 \, \pi\, \as^2\,\rho\, T(\un b),
\ee
with $\rho$ the atomic number density in the nucleus with atomic
number $A$, $T(\un b)$ the nuclear profile function with ${\un b} =
({\un x}_{2} + {\un x}_{2'})/2$ and $\Lambda$ some infrared cutoff.

In the diagram in \fig{em2}B the softer gluon $\# 1$ can not be
emitted off gluon $2'$ in the complex conjugate amplitude: that would
be suppressed due to the inverse ordering rule we derived in
Sect. IIA. Furthermore, if the gluon $2'$ is emitted off the quark
line $0$ the gluon $\# 1$ can not be emitted in the dipole formed by
lines $0$ and $2'$ due to the same inverse ordering rule. Therefore,
if the gluon $2'$ is emitted off the quark line $0$, the gluon $\# 1$
can only be emitted by the dipole $0{\tilde 0}$ in the complex
conjugate amplitude. Therefore, the diagram \fig{em2}B would bring in
a factor of $M_0 ({\un x}_2, {\un x}_0, {\un x}_{\tilde 0}; {\un
k}_1)$ if the gluon $\# 1$ is emitted in the lower ``dipole''.  In the
same case, in the upper ``dipole'', only the dipole $02$ would
interact with the target bringing in a factor of $S_0 ({\un x}_0, {\un
x}_2)$. The diagram in \fig{em2}C can be obtained from \ref{em2}B by
horizontal reflection which can be accomplished by interchanging ${\un
x}_2
\leftrightarrow {\un x}_{2'}$. Finally, a similar line of arguments
shows that the diagram in \fig{em2}D brings in a factor of $M_0 ({\un
x}_0, {\un x}_0, {\un x}_{\tilde 0}; {\un k}_1)$ if the gluon is
emitted in the lower ``dipole''.

Combining the diagrams A-D in \fig{em2} and defining
\be
\bas \, \equiv \, \frac{\as \, N_c}{\pi}
\ee
we write
\ben
\frac{d {\hat \sigma}^{q{\bar q}A\rightarrow q{\bar q}GGX}}{d^2 k_1 \, 
dy_1 \, d^2 k_2 \, dy_2}({\underline x}_{0{\tilde 0}}) \bigg|_{y_2 \gg
y_1} \, = \,
\frac{\bas}{(2 \pi)^3}
\, \int \, d^2 B \, d^2 x_2 \, d^2 x_{2'} \, 
e^{- i {\underline k}_2 \cdot {\underline x}_{22'}} \,
\bigg[ \left( \frac{{\un x}_{20}}{x_{20}^2} - 
\frac{{\un x}_{2{\tilde 0}}}{x_{2{\tilde 0}}^2} \right) \cdot 
\left( \frac{{\un x}_{2'0}}{x_{2'0}^2} - 
\frac{{\un x}_{2'{\tilde 0}}}{x_{2'{\tilde 0}}^2} \right) 
\een
\ben
\times \, M_0 ({\un x}_2, {\un x}_{2'}, {\un x}_{\tilde 0}; {\un k}_1) 
\, S_0 ({\un x}_2, {\un x}_{2'}) - \left( \frac{{\un
x}_{20}}{x_{20}^2} -
\frac{{\un x}_{2{\tilde 0}}}{x_{2{\tilde 0}}^2} \right) \cdot 
\frac{{\un x}_{2'0}}{x_{2'0}^2} \, 
M_0 ({\un x}_2, {\un x}_0, {\un x}_{\tilde 0}; {\un k}_1) \ S_0({\un
x}_0, {\un x}_2) -
\een
\be\label{2Gcl}
- \left( \frac{{\un x}_{2'0}}{x_{2'0}^2} -
\frac{{\un x}_{2'{\tilde 0}}}{x_{2'{\tilde 0}}^2} \right) \cdot 
\frac{{\un x}_{20}}{x_{20}^2} 
\, M_0 ({\un x}_0, {\un x}_{2'}, {\un x}_{\tilde 0}; {\un k}_1) \ 
S_0 ({\un x}_0, {\un x}_{2'}) + \frac{{\un x}_{20}}{x_{20}^2} \cdot
\frac{{\un x}_{2'0}}{x_{2'0}^2}
\, M_0 ({\un x}_0, {\un x}_0, {\un x}_{\tilde 0}; {\un k}_1) 
+ (0 \leftrightarrow {\tilde 0}) \bigg] \bigg\},
\ee
where ${\un B} = ({\un x}_0 + {\un x}_{\tilde 0})/2$ is the impact
parameter of the original dipole $0{\tilde 0}$. The term $(0
\leftrightarrow {\tilde 0})$ implies that we have to add the whole
expression again interchanging $0$ and ${\tilde 0}$ to account for the
emission of gluon $\# 1$ from the top ``dipole''.
\begin{figure}
\begin{center}
\epsfxsize=15cm
\leavevmode
\hbox{\epsffile{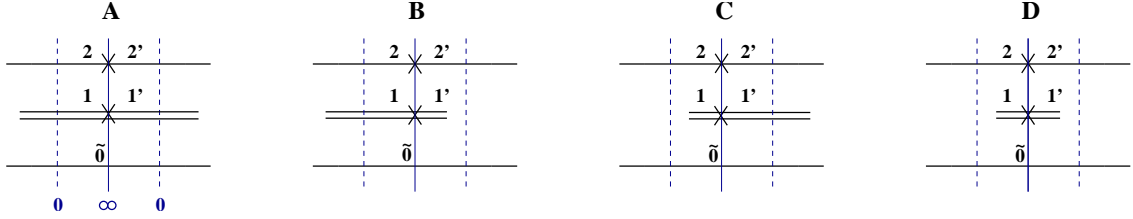}}
\end{center}
\caption{All possible emissions of the softer gluon $\# 1$ by the 
``dipole'' $2,2',{\tilde 0}$.}
\label{em1}
\end{figure}
Now we have to calculate $M_0 ({\un x}_2, {\un x}_{2'}, {\un
x}_{\tilde 0}; {\un k}_1)$. To do that let us consider all possible
emissions of the gluon $\# 1$ in the ``dipole'' $2,2',{\tilde 0}$ as
shown in \fig{em1}. The transverse coordinates of gluon $\# 1$ are
${\un x}_1$ and ${\un x}_{1'}$ to the left and to the right of the cut
correspondingly.

To calculate all the diagrams in \fig{em1} one has to use the rules of
Sect. IIA. Let us illustrate the prescription for calculating these
graphs for the fairly general case when the gluon $\# 1$ is emitted
off gluon $\# 2$ on both sides of the cut. The ``dipole''
$2,2',{\tilde 0}$ in \fig{em1}A would then split into a ``dipole''
$1,1',{\tilde 0}$ and a quadrupole $2,2',1,1'$. The interactions of
line ${\tilde 0}$ with the target cancel via real-virtual
cancellations \cite{km,KMc}, thus reducing the interactions of the
``dipole'' $1,1',{\tilde 0}$ to the interaction of a real dipole
$1,1'$ bringing in a factor of $S_0 ({\un x}_1,{\un x}_{1'} )$. The
interaction of the quadrupole $2,2',1,1'$ brings in a factor which we
will denote $Q_0 ({\un x}_2, {\un x}_{2'}, {\un x}_1, {\un
x}_{1'})$. In the quasi-classical approximation of
McLerran-Venugopalan model this $S$-matrix of the quadrupole
interaction with the target $Q_0$ is calculated in the Appendix A
yielding (cf. \cite{nszz})
\ben
Q_0 ({\un x}_2, {\un x}_{2'}, {\un x}_1, {\un x}_{1'}) \, = \,
e^{- [x_{21}^2 \ln (1/x_{21} \Lambda)+
x_{2'1'}^2 \ln (1/x_{2'1'} \Lambda)] Q_{s0}^2 /4} 
\een
\ben
+ \, \frac{x_{22'}^2 \ln (1/x_{22'} \Lambda)+ x_{11'}^2 \ln
(1/x_{11'} \Lambda) - x_{21'}^2 \ln (1/x_{21'} \Lambda) - x_{2'1}^2 \ln
(1/x_{2'1} \Lambda)}{x_{21}^2 \ln (1/x_{21} \Lambda)+ x_{2'1'}^2 \ln
(1/x_{2'1'} \Lambda) - x_{22'}^2 \ln (1/x_{22'} \Lambda) - x_{11'}^2 \ln
(1/x_{11'} \Lambda)}
\een
\be\label{Q0}
\times \, \left( e^{- [x_{21}^2 \ln (1/x_{21} \Lambda)+
x_{2'1'}^2 \ln (1/x_{2'1'} \Lambda)] Q_{s0}^2 /4} - e^{- [x_{11'}^2
\ln (1/x_{11'} \Lambda)+ x_{22'}^2 \ln (1/x_{22'} \Lambda)] Q_{s0}^2
/4} \right).
\ee
As once can see from \eq{Q0}, $Q_0 ({\un x}_2, {\un x}_{2}, {\un x}_1,
{\un x}_{1}) = 1$ which is what one would expect due to real-virtual
cancellations \cite{km,KMc,kt}. One can also check that
\be\label{test0}
Q_0 ({\un x}_2, {\un x}_{2'}, {\un x}_1, {\un x}_{1}) \, = \, e^{-
x_{22'}^2 \ln (1/x_{22'} \Lambda) Q_{s0}^2 /4}
\ee
corresponding to the $S$-matrix of the interaction of the dipole $22'$
with the target. (Interactions with line $1$ cancel again due to
real-virtual cancellations if we put ${\un x}_1 = {\un x}_{1'}$.)

Now let us evaluate the graph in \fig{em1}B for the same case of gluon
$\# 1$ being emitted off gluon $\# 2$ on both sides of the cut. In the
top ``dipole'' interactions can only take place to the left of the cut
giving a factor of $S_0 ({\un x}_2,{\un x}_1 )$. In the bottom
``dipole'' the interactions with line $\# {\tilde 0}$ cancel, leaving
only the dipole $12'$ to interact with the target which brings in a
factor of $S_0 ({\un x}_1,{\un x}_{2'} )$. The diagram in
\fig{em1}C is evaluated in a similar way yielding a factor of 
$S_0 ({\un x}_{2'},{\un x}_{1'} ) \, S_0 ({\un x}_2,{\un
x}_{1'})$. Finally, in the diagram in \fig{em1}D only the lower dipole
can interact with the target giving a factor of $S_0 ({\un x}_2,{\un
x}_{2'})$.

Combining the factors calculated above for the diagrams A-D in
\fig{em1}, putting in the contribution of gluon emission and summing over 
all possible connections of gluon $\# 1$ to the lines $2, 2'$ and $1$
we obtain
\ben
M_0 ({\un x}_2, {\un x}_{2'}, {\un x}_{\tilde 0} , {\un k}_1) \, = \,
\frac{\bas}{(2 \pi)^3} \, \int \, d^2 x_1 \, d^2
x_{1'} \, e^{- i {\un k}_1 \cdot {\un x}_{11'}} \, \bigg\{ 
\frac{{\un x}_{12}}{x_{12}^2} \cdot 
\frac{{\un x}_{1'2'}}{x_{1'2'}^2} \ \bigg[ Q_0 ({\un x}_2, 
{\un x}_{2'}, {\un x}_1, {\un x}_{1'}) \, S_0 ({\un x}_1, {\un
x}_{1'}) 
\een
\ben
+ S_0 ({\un x}_2, {\un x}_{2'}) - S_0 ({\un x}_2, {\un x}_{1}) \, 
S_0 ({\un x}_1, {\un x}_{2'})  - S_0 ({\un x}_{2'}, {\un x}_{1'}) \, 
S_0 ({\un x}_2, {\un x}_{1'})  \bigg]
+ \frac{{\un x}_{1{\tilde 0}}}{x_{1{\tilde 0}}^2} \cdot 
\frac{{\un x}_{1'{\tilde 0}}}{x_{1'{\tilde 0}}^2} 
\bigg[ Q_0 ({\un x}_2, {\un x}_{2'}, {\un x}_1, {\un x}_{1'})  
\een
\ben
\times S_0 ({\un x}_1, {\un x}_{1'}) + S_0 ({\un x}_2, {\un x}_{2'}) 
- Q_0 ({\un x}_2, {\un x}_{2'}, {\un x}_1, {\un x}_{\tilde 0}) \,
S_0 ({\un x}_1, {\un x}_{\tilde 0}) - Q_0 ({\un x}_2, {\un x}_{2'}, {\un
x}_{\tilde 0}, {\un x}_{1'}) \,  S_0 ({\un x}_{\tilde 0}, {\un x}_{1'})
\bigg]
\een
\ben
- \frac{{\un x}_{12}}{x_{12}^2} \cdot 
\frac{{\un x}_{1'{\tilde 0}}}{x_{1'{\tilde 0}}^2} \, 
\bigg[ Q_0 ({\un x}_2, {\un x}_{2'}, {\un x}_1, {\un x}_{1'}) \, 
S_0 ({\un x}_1, {\un x}_{1'})  + S_0 ({\un x}_2, {\un x}_{\tilde 0}) 
\, S_0 ({\un x}_{2'}, {\un x}_{\tilde 0})  
- Q_0 ({\un x}_2, {\un x}_{2'}, {\un x}_1, {\un x}_{\tilde 0}) \,
S_0 ({\un x}_1, {\un x}_{\tilde 0})
\een
\ben
- S_0 ({\un x}_2, {\un x}_{1'}) \,
S_0 ({\un x}_{2'}, {\un x}_{1'}) \bigg] 
- \frac{{\un x}_{1{\tilde 0}}}{x_{1{\tilde 0}}^2} \cdot \frac{{\un
x}_{1'2'}}{x_{1'2'}^2} \, \bigg[ Q_0 ({\un x}_2, {\un x}_{2'}, {\un
x}_1, {\un x}_{1'}) \, S_0 ({\un x}_1, {\un x}_{1'})  + 
S_0 ({\un x}_2, {\un x}_{\tilde 0}) \, 
S_0 ({\un x}_{2'}, {\un x}_{\tilde 0}) 
\een
\be\label{M0}
-  S_0 ({\un x}_2, {\un x}_1) \, S_0 ({\un x}_{2'}, {\un x}_1)
- Q_0 ({\un x}_2, {\un x}_{2'}, {\un x}_{\tilde 0}, {\un x}_{1'}) \,
S_0 ({\un x}_{1'}, {\un x}_{\tilde 0}) \bigg] \bigg\}
\ee
Eqs. (\ref{disincl}) and (\ref{2Gcl}), together with Eqs. (\ref{M0}),
(\ref{Q0}), (\ref{S0}) and (\ref{N0}) give us the two-gluon inclusive
production cross section for DIS on a nucleus including all the
quasi-classical multiple rescatterings in the large-$N_c$
approximation. It is the main result of this Section.

\section{Two-Gluon Production Including Quantum Evolution}

In this Section our goal is to include the effects of non-linear
small-$x$ quantum evolution of \cite{bk} into the quasi-classical
expression (\ref{2Gcl}) for inclusive two-gluon production cross
section. We will begin by reviewing the non-linear evolution equation
and its application to single inclusive gluon production. We will
proceed by deriving the expression generalizing \eq{2Gcl} by including
the non-linear evolution \cite{bk} in it. We will conclude by
verifying that the obtained expression matches onto the standard
$k_T$-factorization result \cite{lr} at the leading twist level.

\subsection{Brief Review of Small-$x$ Evolution and Single Gluon Production}

To include the effects of small-$x$ evolution in the dipole $S$-matrix
one first defines the $S$ matrix for a quark dipole $0{\tilde 0}$
having rapidity $Y$ with respect to the target as
\be\label{S}
S ({\un x}_{0}, {\un x}_{\tilde 0}, Y) = 1 - N ({\un x}_{0}, {\un
x}_{\tilde 0}, Y),
\ee
where the forward scattering amplitude has to be determined from the
non-linear evolution equation \cite{bk}
\ben
  N({\underline x}_{0}, {\un x}_{\tilde 0}, Y) = N_0 ({\underline
  x}_{0}, {\un x}_{\tilde 0}) \, e^{- 2 \, \bas \, \ln
  \left( \frac{x_{0{\tilde 0}}}{\rho} \right) Y} + \frac{\bas}{2 \, \pi}
  \int_0^Y d y \ e^{ - 2\, \bas \, \ln \left(
  \frac{x_{0{\tilde 0}}}{\rho} \right) (Y - y)}
\een
\be\label{eqN}
\times \int_\rho d^2 x_2 \frac{x_{0{\tilde 0}}^2}{x_{20}^2 
x_{2{\tilde 0}}^2} \, [ N ({\underline x}_{0},{\un x}_2, y) + N
({\underline x}_{2},{\un x}_{\tilde 0}, y) - 
N ({\underline x}_{0},{\un x}_2, y) \, N
({\underline x}_{2},{\un x}_{\tilde 0}, y) ] ,
\ee
with the initial condition given by \eq{N0} and $\rho$ being an
ultraviolet cutoff \cite{dip}. The evolution equation (\ref{eqN})
resums all powers of leading logarithms of center of mass energy $\as
Y$ and all multiple interactions with the target, which bring in
powers of $\as^2 A^{1/3}$ with $A$ the atomic number of the nucleus
\cite{kjkmw,km,KMc,yuri1}.

\eq{eqN} is derived in \cite{bk} by resumming a cascade of gluons in 
the incoming $q\bar q$ wave function, which in the large-$N_c$ limit
turns into a cascade of color dipoles. The emissions are similar to
the ones we considered in Sect. IIA for the early times preceding the
interaction. The difference is that in \cite{bk} one resums emissions
to all orders, without limiting oneself to just two gluons. At the
leading logarithmic level the contribution of this gluonic (dipole)
cascade to the $S$-matrix of the dipole-nucleus scattering is given by
the solution of \eq{eqN}. However, one can use this cascade to
construct other useful observables.

In \cite{kt} it was shown that inclusive gluon production cross
section in a dipole-nucleus scattering is given by the following
formula:
\be\label{1Gincl}
\frac{d \sigma^{q {\bar q} A \rightarrow q{\bar q}GX}}{d^2 k \ dy} 
({\un x}_{0 {\tilde 0}})
\ = \ \int \, d^2 B \ 
n_1 ({\underline x}_0, {\underline x}_{\tilde 0}, Y; {\un x}_1, {\un
x}_2, y) \, d^2 x_1 \, d^2 x_2 \ s ({\un x}_1, {\un x}_2, {\un k},
y),
\ee
where we defined
\ben
s ({\un x}_1, {\un x}_2, {\un k}, y) \, \equiv \, \frac{\bas}{(2
\pi)^3} \, \int \, d^2 z_1 \, d^2 z_2 \,
e^{- i {\underline k} \cdot ({\underline z}_1 - {\underline z}_2 )} \,
\sum_{i,j = 1}^2 (-1)^{i+j} \,
\frac{{\underline z}_1 - {\underline x}_i}{|{\underline z}_1 - 
{\underline x}_i|^2} \cdot \frac{{\underline z}_2 - {\underline
x}_j}{|{\underline z}_2 - {\underline x}_j|^2} 
\een
\be\label{sdef}
\times \, \left[ N_G \left({\underline z}_1, {\underline x}_j, y \right) 
+ N_G \left({\underline z}_2 , {\underline x}_i, y \right) 
- N_G \left({\underline z}_1 , {\underline z}_2, y \right)
- N_G \left({\underline x}_i , {\underline x}_j, y \right) \right]
\ee
in terms of the forward scattering amplitude of the adjoint (gluon)
dipole, which in the large $N_c$ limit can be easily expressed in
terms of the forward amplitude of the fundamental (quark) dipole
\be\label{ngn}
N_G ({\underline x}_0, {\underline x}_1, y) \, = \, 2 \, N
({\underline x}_0, {\underline x}_1, y) - N^2 ({\underline x}_0,
{\underline x}_1, y).
\ee

In \eq{1Gincl} the quantity $n_1 ({\underline x}_0, {\underline
x}_{\tilde 0}, Y; {\un x}_1, {\un x}_2, y)$ has the meaning of the
probability of finding a dipole $12$ at rapidity $y$ in the original
dipole $0{\tilde 0}$ having rapidity $Y$ \cite{dip}. It obeys the
following equation \cite{dip}
\ben
  n_1 ({\underline x}_{0}, {\un x}_{\tilde 0}, Y; {\un x}_1, {\un
  x}_{\tilde 1}, y) = \delta^2 ({\underline x}_{0} - {\un x}_1) \, 
\delta ( {\un x}_{\tilde 0} - {\un x}_{\tilde 1}) \, e^{- 2
  \, \bas \, \ln \left( \frac{x_{0{\tilde 0}}}{\rho} \right) (Y-y)} +
  \frac{\bas}{2 \, \pi} \int_{y}^Y d y' \, e^{ - 2\, \bas \, \ln \left(
  \frac{x_{0{\tilde 0}}}{\rho} \right) (Y - y')}
\een
\be\label{n1}
\times \int_\rho d^2 x_2 \frac{x_{0{\tilde 0}}^2}{x_{20}^2 
x_{2{\tilde 0}}^2} \, [ n_1 ({\underline x}_{0},{\un x}_2, y'; {\un
x}_1, {\un x}_{\tilde 1}, y) + n_1 ({\underline x}_{2},{\un
x}_{\tilde 0}, y'; {\un x}_1, {\un x}_{\tilde 1}, y) ],
\ee
which is the linear part of the dipole evolution equation (\ref{eqN})
equivalent to the BFKL equation \cite{bfkl}.

The quantity $s ({\un x}_1, {\un x}_2, {\un k}, y)$ in \eq{1Gincl} is
the cross section for single gluon production by the dipole $12$
scattering on a nucleus at rapidity $y$ with the emitted gluon being
the first (hardest) gluon in the gluonic (dipole) cascade developed by
the incoming dipole $12$. Than \eq{1Gincl} has a simple physical
meaning: it convolutes the probability of finding a dipole in the
initial onium wave function which would emit the gluon with the
probability of the gluon emission by this dipole.

\begin{figure}
\begin{center}
\epsfxsize=8cm
\leavevmode
\hbox{\epsffile{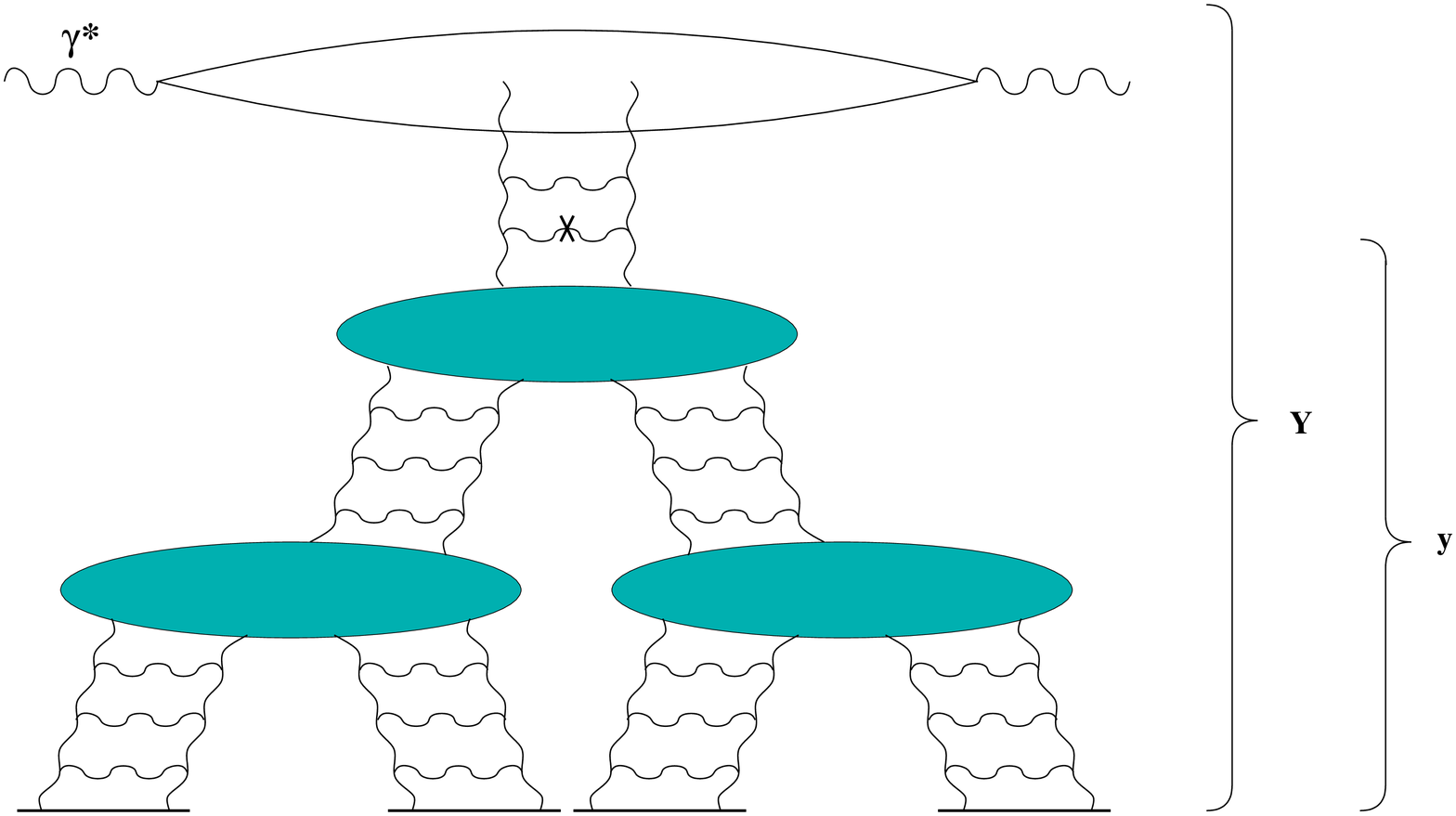}}
\end{center}
\caption{Feynman diagram corresponding to single gluon production 
cross section given by \eq{1Gincl}. Emitted gluon is denoted by the
cross.  }
\label{fan1}
\end{figure}

To recover the quasi-classical result for single gluon production
\cite{km} one has to put $Y=y=0$ on the right hand side of \eq{1Gincl}. 
That would effectively turn off the quantum evolution giving
\ben
\frac{d \sigma^{q {\bar q} A \rightarrow q{\bar q}GX}}{d^2 k \ dy} 
({\un x}_{0 {\tilde 0}})
\,  =  \, \int \, d^2 B \ 
s ({\un x}_0, {\un x}_{\tilde 0}, {\un k}, 0) \,
=  \, \int \, d^2 B \,
d^2 z_1 \, d^2 z_2 \, e^{- i {\underline k} \cdot ({\underline z}_1 -
{\underline z}_2 )} 
\een
\ben
\times \, \sum_{i,j = 1}^2 (-1)^{i+j} \,
\frac{{\underline z}_1 - {\underline x}_i}{|{\underline z}_1 - 
{\underline x}_i|^2} \cdot \frac{{\underline z}_2 - {\underline
x}_j}{|{\underline z}_2 - {\underline x}_j|^2} \, \left( e^{- ({\un x}_i -
{\un x}_j)^2 Q_{s0}^2\ln(1/|{\un x}_i-{\un x}_j| \Lambda) /2 } - e^{-
({\un z}_1 - {\un x}_j)^2 Q_{s0}^2\ln(1/|{\un z}_1 - {\un x}_j|
\Lambda) /2 } \right.
\een
\be
\left. - e^{- ({\un z}_2 - {\un x}_i)^2 Q_{s0}^2\ln(1/|{\un z}_2 
- {\un x}_i|\Lambda) /2 } + e^{-
({\un z}_1 - {\un z}_2)^2 Q_{s0}^2\ln(1/|{\un z}_1-{\un z}_2| \Lambda)
/2 } \right) \label{1Gcl}
\ee
which is the quasi-classical gluon production cross section found in
\cite{km,kst,dmc,yuri1}.

If the evolution equation (\ref{eqN}) is pictured as resumming the
so-called ``fan'' diagrams in dipole-nucleus scattering \cite{glr,mq},
than the single inclusive gluon production cross section would
correspond to diagrams like the one shown in \fig{fan1}. There the
produced gluon, which is denoted by the cross, can be emitted only
from the top ladder in the diagram. As it turned out, emissions from
all other (lower) ladders cancel \cite{kt}, in agreement with
expectations of the AGK cutting rules \cite{agk} (see also
\cite{braun}). Thus the evolution between the projectile and the
produced gluon is just a linear BFKL evolution, as we can see in
\eq{n1}. The evolution between the produced gluon and the target is
the full non-linear evolution given by \eq{eqN}, as can bee seen from
\eq{1Gincl}.

Before concluding the~subsection let us define another useful
quantity. Following \cite{dip} let 
\ben
n_2 ({\un x}_0, {\un x}_{\tilde 0}, Y; {\un x}_1 , {\un x}_{\tilde 1},
y_1, {\un x}_2, {\un x}_{\tilde 2}, y_2)
\een
be the probability of finding dipoles $1{\tilde 1}$ and $2{\tilde 2}$
with rapidities $y_1$ and $y_2$ correspondingly in the original dipole
$0{\tilde 0}$ having rapidity $Y$. This quantity obeys the following
evolution equation
\cite{dip}
\ben
n_2 ({\un x}_0, {\un x}_{\tilde 0}, Y; {\un x}_1 , {\un x}_{\tilde 1},
y_1, {\un x}_2, {\un x}_{\tilde 2}, y_2) \, = \, \frac{\bas}{2 \, \pi} \,
\int_{\mbox{max} \{ y_1, y_2 \}}^Y d y \ e^{ - 2 \, \bas \, \ln \left(
  \frac{x_{0{\tilde 0}}}{\rho} \right) (Y - y)} \, \int_\rho d^2 x_3
  \frac{x_{0{\tilde 0}}^2}{x_{30}^2 x_{3{\tilde 0}}^2}
\een
\ben
\times  \, \left[ n_1 ({\un
  x}_0, {\un x}_3, y; {\un x}_1 , {\un x}_{\tilde 1}, y_1) \, n_1
  ({\un x}_3, {\un x}_{\tilde 0}, y; {\un x}_2 , {\un x}_{\tilde 2},
  y_2) + n_1 ({\un
  x}_0, {\un x}_3, y; {\un x}_2 , {\un x}_{\tilde 2}, y_2) \, n_1
  ({\un x}_3, {\un x}_{\tilde 0}, y; {\un x}_1 , {\un x}_{\tilde 1},
  y_1) +  \right.
\een
\be\label{n2}
+ \left. n_2 ({\un x}_0, {\un x}_3, y; {\un x}_1 , {\un x}_{\tilde 1},
y_1, {\un x}_2 , {\un x}_{\tilde 2}, y_2) + n_2 ({\un x}_3, {\un
x}_{\tilde 0}, y; {\un x}_1 , {\un x}_{\tilde 1}, y_1, {\un x}_2 ,
{\un x}_{\tilde 2}, y_2)\right]
\ee
which is linear and can be solved after one finds $n_1$ from
\eq{n1}.

\subsection{Two-Gluon Inclusive Cross Section with Quantum Evolution}

Now we have all the essential ingredients necessary to include quantum
evolution effects into \eq{2Gcl}. Similar to the analysis carried out
in \cite{kt} we will separate all the gluons into the ones which are
harder (have higher rapidity with respect to the target) than the
harder of the two gluons with rapidities $y_1$ and $y_2$ that are
going to be produced and into the ones which are softer (have lower
rapidity) than the gluon $y_2$ ($y_2 \gg y_1$).

Similarly to the analysis of Sect. IIIA in \cite{kt}, one can easily
conclude that all of the harder gluons can be emitted only at early
($\tau < 0$) times both in the amplitude and in the complex conjugate
amplitude. Due to the ordering rule from Sect. IIA of this paper, this
implies that these harder gluons have to be emitted before gluons $\#
2$ and $\# 1$. Therefore, we have to distinguish two important cases:

\begin{enumerate}

\item[{\bf A.}] The gluons $\# 2$ and $\# 1$ are emitted in two 
different dipoles created by the evolution due to emission of gluons
which are harder than either gluon $\# 2$ or gluon $\# 1$.

\item[{\bf B.}] The gluon $\# 2$ is emitted in a dipole created by the 
evolution consisting of emissions of harder gluons. The gluon $\# 1$
is emitted either by one of the ``dipoles'' adjacent to gluon $\# 2$
(as was studied in \fig{em1}) or in a dipole generated by evolution
inside one of these adjacent ``dipoles''.

\end{enumerate}

Case A is relatively straightforward. Quantum evolution creating two
dipoles of given sizes and rapidities at times $\tau < 0$ is included
in the quantity $n_2$ from \eq{n2}. Emission of each of the gluons $\#
1$ and $\# 2$ in two independent dipoles is equivalent to the same
problem of a single inclusive gluon emission in a dipole-nucleus
collision as considered in Sect. IIIA and is described by the quantity
$s$ from \eq{sdef}, which also includes all the successive evolution
generated through emissions of gluons softer than either $\# 1$ or $\#
2$ \cite{kt}. Therefore the contribution of case A to double gluon
production can be written as
\be\label{Apiece}
\int d^2 B \, 
n_2 ({\un x}_0, {\un x}_{\tilde 0}, Y; {\un x}_1 , {\un x}_{\tilde 1},
y_1, {\un x}_2, {\un x}_{\tilde 2}, y_2) \, d^2 x_1 \, d^2 x_{\tilde
1} \, d^2 x_2 \, d^2 x_{\tilde 2} \, s ({\un x}_1, {\un x}_{\tilde 1},
{\un k}_1, y_1) \, s ({\un x}_{2}, {\un x}_{\tilde 2}, {\un k}_2,
y_2)
\ee
with ${\un B} = ({\un x}_0 + {\un x}_{\tilde 0})/2$ as before.

\begin{figure}
\begin{center}
\epsfxsize=12cm
\leavevmode
\hbox{\epsffile{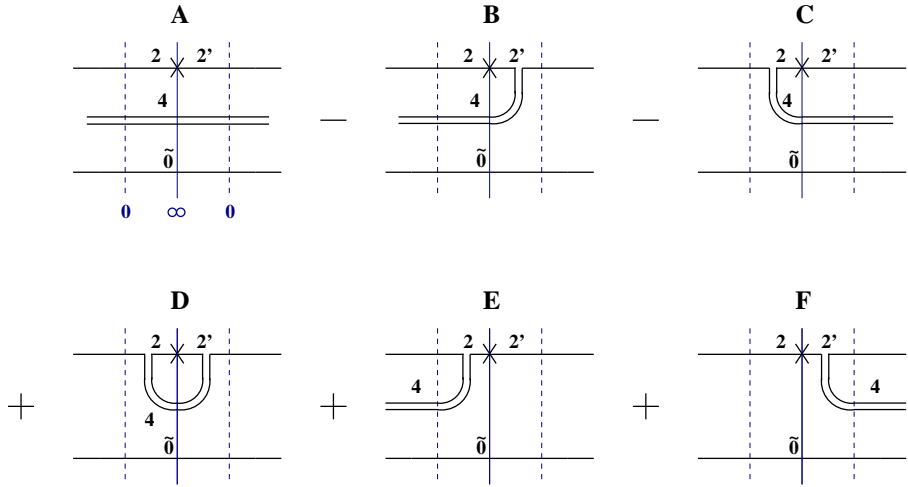}}
\end{center}
\caption{Diagrams describing one step of evolution for $M$.}
\label{mev}
\end{figure}

Contribution of case B is somewhat more complicated. The probability
of finding an early time ($\tau < 0$) dipole in the original onium in
which gluon $\# 2$ is emitted is described by the quantity $n_1$ from
\eq{n1}. Emission of gluon $\# 2$ is then described by the diagrams of
\fig{em2} and, equivalently, by \eq{2Gcl}. The only difference is that
now we have to include quantum evolution in the quantity $M_0$ and in
the $S$-matrix $S_0$. The inclusion of evolution into the $S$-matrix
$S_0$ is accomplished in Eqs. (\ref{S}) and (\ref{eqN}). The inclusion
of evolution into the quantity $M_0$ requires a separate diagrammatic
analysis, shown in \fig{mev}.

Let us first define a quantity $M ({\un x}_2, {\un x}_{2'}, {\un
x}_{\tilde 0}, Y; {\un k}_1, y_1)$ which by analogy with $M_0$ has a
physical meaning of an inclusive cross section of producing a gluon
with transverse momentum ${\un k}_1$ and rapidity $y_1$ in the
``dipole'' $2, 2', {\tilde 0}$ having rapidity $Y$. To write down an
evolution equation for $M ({\un x}_2, {\un x}_{2'}, {\un x}_{\tilde
0}, Y; {\un k}_1, y_1)$ one has to analyze a single step of small-$x$
evolution for this quantity. All the important gluon emissions in the
``dipole'' $2, 2', {\tilde 0}$ are shown in \fig{mev}. We start the
analysis with the diagram in \fig{mev}A. Emitting the gluon $\# 4$
splits the original ``dipole'' $2, 2', {\tilde 0}$ into a ``dipole''
$2, 2', 4$ and a dipole $4 {\tilde 0}$. Than the gluon $\# 1$ can be
emitted in the ``dipole'' $2, 2', 4$, which would bring in a factor of
\ben
M ({\un x}_2, {\un x}_{2'}, {\un x}_4, y; {\un k}_1, y_1)
\een
with $y$ the rapidity of gluon $\# 4$. In this case all interactions
in dipole $4 {\tilde 0}$ cancel. Alternatively the gluon $\# 1$ can be
emitted in the dipole $4 {\tilde 0}$, which would bring in the
familiar factor of
\ben
\int d^2 x_a d^2 x_b \, n_1 ({\un x}_{4}, {\un x}_{\tilde 0}, y;
{\un x}_a, {\un x}_b, y_1) \, s ({\un x}_a, {\un x}_b, {\un k}_1, y_1)
\een
from \eq{1Gincl} describing a single inclusive gluon production in a
dipole-nucleus scattering. In this second case interactions of the
``dipole'' $2, 2', 4$ with the target would not completely
cancel. Instead they would bring in a factor of $S ({\un x}_2, {\un
x}_{2'}, y) = 1 - N ({\un x}_2, {\un x}_{2'}, y)$ corresponding to
interaction of dipole $22'$ with the target.

In the diagram shown in \fig{mev}B the interaction of gluon $\# 4$
with the line $\tilde 0$ to the right of the cut gets canceled by the
diagram similar to the one in \fig{mev}E but with the gluon $\# 4$
connecting to the line $\tilde 0$ instead of line $2$ to the right of
the cut \cite{CM}. Therefore, gluon $\# 4$ interacts only with line
$2'$ on the right hand side of \fig{mev}B and only with line $2$ on
the right hand side of \fig{mev}E. Due to the inverse ordering rule
for late time emissions from Sect. IIA the gluon $\# 1$ can not be
emitted in the dipole $24$ in graph B of \fig{mev}. (In \fig{mev}E the
dipole $24$ is not present in the final state and can not contribute
to gluon production.) Therefore, in both graphs B and E the dipole
$24$ brings in only a factor of $S ({\un x}_2, {\un x}_4, y) = 1 - N
({\un x}_2, {\un x}_4, y)$ into the evolution equation we are
constructing. The other ``dipole'' $4, 2', {\tilde 0}$ would than
bring in a factor of
\ben
M ({\un x}_4, {\un x}_{2'}, {\un x}_{\tilde 0}, y; {\un k}_1, y_1).
\een
Diagrams C and F can be obtained from B and E by interchanging $2
\leftrightarrow 2'$. Finally, in the diagram D interactions of gluon 
$\# 4$ with the line ${\tilde 0}$ cancel due to real-virtual
cancellations of \cite{CM} that we have just employed in graphs B and
E. The remaining interaction with lines $2$ and $2'$ shown in
\fig{mev}D does not split the ``dipole'' $2,2',{\tilde 0}$. Due to 
inverse ordering rule of Sect. IIA there will be no softer gluon
emissions in ``dipole'' $2,2'4$, such that all subsequent evolution
will take place only in ``dipole'' $2,2',{\tilde 0}$.  Therefore,
diagram D contributes only to virtual corrections, along with the
usual virtual corrections at $\tau < 0$ in dipoles $2 {\tilde 0}$ and
$2' {\tilde 0}$.

Combining the contributions of all diagrams in \fig{mev} we write the
following evolution equation:
\ben
M ({\un x}_2, {\un x}_{2'}, {\un x}_{\tilde 0}, Y; {\un k}_1, y_1) \,
= \, e^{- \bas \, \ln \left( \frac{x_{2{\tilde 0}} x_{2'{\tilde 0}}
x_{22'}}{\rho^3} \right) \, (Y-y_1)} \, d({\un x}_2, {\un x}_{2'},
{\un x}_{\tilde 0}, {\un k}_1, y_1) +
\een
\ben
+ \frac{\bas}{2 \pi} \, \int d^2 x_4 \int_{y_1}^Y dy \, e^{- \bas \,
\ln \left( \frac{x_{2{\tilde 0}} x_{2'{\tilde 0}} x_{22'}}{\rho^3}
\right) \, (Y-y)} \, \bigg\{
\left( \frac{{\un x}_{42}}{x_{42}^2} - 
\frac{{\un x}_{4{\tilde 0}}}{x_{4{\tilde 0}}^2} \right) \cdot 
\left( \frac{{\un x}_{42'}}{x_{42'}^2} - 
\frac{{\un x}_{4{\tilde 0}}}{x_{4{\tilde 0}}^2} \right) 
\een
\ben
\times 
\bigg[ M ({\un x}_2, {\un x}_{2'}, {\un x}_4, y; {\un k}_1, y_1) 
+ \int d^2 x_a d^2 x_b \, n_1 ({\un x}_4, {\un x}_{\tilde 0}, y;
{\un x}_a, {\un x}_b, y_1) \, s ({\un x}_a, {\un x}_b, {\un k}_1, y_1)
\, [1 - N ({\un x}_2, {\un x}_{2'}, y)] \bigg] - 
\een
\ben
- \left( \frac{{\un x}_{42}}{x_{42}^2} - 
\frac{{\un x}_{4{\tilde 0}}}{x_{4{\tilde 0}}^2} \right) \cdot 
\left( \frac{{\un x}_{42'}}{x_{42'}^2} - 
\frac{{\un x}_{42}}{x_{42}^2} \right) \, 
M ({\un x}_4, {\un x}_{2'}, {\un x}_{\tilde 0}, y; {\un k}_1, y_1) 
\, [1 - N ({\un x}_2, {\un x}_4, y)] - 
\een
\be\label{Mev}
- \left( \frac{{\un x}_{42}}{x_{42}^2} - 
\frac{{\un x}_{42'}}{x_{42'}^2} \right) \cdot 
\left( \frac{{\un x}_{42'}}{x_{42'}^2} - 
\frac{{\un x}_{4{\tilde 0}}}{x_{4{\tilde 0}}^2} \right) \, 
M ({\un x}_2, {\un x}_4, {\un x}_{\tilde 0}, y; {\un k}_1, y_1) 
\, [1 - N ({\un x}_{2'}, {\un x}_4, y)] \bigg\} .
\ee
The only thing left to do to complete our analysis is to determine the
initial condition for the evolution equation (\ref{Mev}), which we
denoted $d({\un x}_2, {\un x}_{2'}, {\un x}_{\tilde 0}, {\un k}_1,
y_1)$. This quantity is the gluon production cross section in the
scattering of a ``dipole'' $2,2',{\tilde 0}$ on a nucleus with the
small-$x$ quantum evolution included, in which the emitted gluon
(${\un k}_1, y_1$) is the first (hardest) gluon in the gluonic cascade
resumming the quantum evolution of \eq{eqN}. Since the emission
diagrams are the same as in \fig{em1}, the quantity $d({\un x}_2, {\un
x}_{2'}, {\un x}_{\tilde 0}, {\un k}_1, y_1)$ should be given by the
expression similar to \eq{M0}, where the dipole and quadrupole
$S$-matrices $S_0$ and $Q_0$ have to be replaced by their evolved
values. We, therefore, write
\ben
d ({\un x}_2, {\un x}_{2'}, {\un x}_{\tilde 0} , {\un k}_1, y_1) \, = \,
\frac{\bas}{(2 \pi)^3} \, \int \, d^2 x_1 \, d^2
x_{1'} \, e^{- i {\un k}_1 \cdot {\un x}_{11'}} \, \bigg\{ 
\frac{{\un x}_{12}}{x_{12}^2} \cdot 
\frac{{\un x}_{1'2'}}{x_{1'2'}^2} \ \bigg[ Q ({\un x}_2, 
{\un x}_{2'}, {\un x}_1, {\un x}_{1'}, y_1) \, S ({\un x}_1, {\un
x}_{1'}, y_1) 
\een
\ben
+ S ({\un x}_2, {\un x}_{2'}, y_1) - S ({\un x}_2, {\un x}_{1}, y_1) \, 
S ({\un x}_1, {\un x}_{2'}, y_1)  - S ({\un x}_{2'}, {\un x}_{1'}, y_1) \, 
S ({\un x}_2, {\un x}_{1'}, y_1)  \bigg]
+ \frac{{\un x}_{1{\tilde 0}}}{x_{1{\tilde 0}}^2} \cdot 
\frac{{\un x}_{1'{\tilde 0}}}{x_{1'{\tilde 0}}^2} 
\bigg[ Q ({\un x}_2, {\un x}_{2'}, {\un x}_1, {\un x}_{1'}, y_1)  
\een
\ben
\times S ({\un x}_1, {\un x}_{1'}, y_1) + S ({\un x}_2, {\un x}_{2'}, y_1) 
- Q ({\un x}_2, {\un x}_{2'}, {\un x}_1, {\un x}_{\tilde 0}, y_1) \, S
({\un x}_1, {\un x}_{\tilde 0}, y_1) - Q ({\un x}_2, {\un x}_{2'},
{\un x}_{\tilde 0}, {\un x}_{1'}, y_1) S ({\un x}_{\tilde 0}, {\un
x}_{1'}, y_1)
\bigg]
\een
\ben
 - \frac{{\un x}_{12}}{x_{12}^2} \cdot 
\frac{{\un x}_{1'{\tilde 0}}}{x_{1'{\tilde 0}}^2} \, 
\bigg[ Q ({\un x}_2, {\un x}_{2'}, {\un x}_1, {\un x}_{1'}, y_1) \, 
S ({\un x}_1, {\un x}_{1'}, y_1)  + S ({\un x}_2, {\un x}_{\tilde 0}, y_1) 
\, S ({\un x}_{2'}, {\un x}_{\tilde 0}, y_1)  
- Q ({\un x}_2, {\un x}_{2'}, {\un x}_1, {\un x}_{\tilde 0}, y_1) 
\een
\ben
\times S ({\un x}_1, {\un x}_{\tilde 0}, y_1) 
- S ({\un x}_2, {\un x}_{1'}, y_1) \,
S ({\un x}_{2'}, {\un x}_{1'}, y_1) \bigg] 
- \frac{{\un x}_{1{\tilde 0}}}{x_{1{\tilde 0}}^2} \cdot \frac{{\un
x}_{1'2'}}{x_{1'2'}^2} \, \bigg[ Q ({\un x}_2, {\un x}_{2'}, {\un
x}_1, {\un x}_{1'}, y_1) \, S ({\un x}_1, {\un x}_{1'}, y_1) 
\een
\be\label{d}
+ S ({\un x}_2, {\un x}_{\tilde 0}, y_1) \, 
S ({\un x}_{2'}, {\un x}_{\tilde 0}, y_1) 
-  S ({\un x}_2, {\un x}_1, y_1) \, S ({\un x}_{2'}, {\un x}_1, y_1)
- Q ({\un x}_2, {\un x}_{2'}, {\un x}_{\tilde 0}, {\un x}_{1'}, y_1) \,
S ({\un x}_{1'}, {\un x}_{\tilde 0}, y_1) \bigg] \bigg\}.
\ee
Indeed $S ({\un x}_1, {\un x}_{1'}, y_1)$ in \eq{d} is given by
Eqs. (\ref{S}) and (\ref{eqN}). The reason why inclusion of evolution
just corresponds to replacing the Glauber-Mueller expression
(\ref{S0}) for $S_0$ by the fully evolved \eq{S} has been discussed
before in \cite{kt}. It was observed there that real-virtual
cancellations for the gluon emissions contributing to the dipole
evolution discussed in \cite{CM} act very much like the real-virtual
cancellations for Glauber-Mueller multiple rescatterings
\cite{km}. Namely, if interactions of exchanged Coulomb gluons with a
given quark line cancel in the multiple rescattering (Glauber-Mueller)
picture, than emissions of an $s$-channel gluon by the same quark line
at early and late times on both sides of the cut would also cancel
\cite{CM}. One can show that interactions with quark and gluon lines
that contribute in multiple rescattering case would also contribute in
the case of evolution.  In the end one concludes that inclusion of
quantum evolution can be accomplished by replacing $S_0$ from
Eqs. (\ref{S0}) and (\ref{N0}) by $S$ from Eqs. (\ref{S}) and
(\ref{eqN}) \cite{kt}.

To calculate $Q ({\un x}_2, {\un x}_{2'}, {\un x}_1, {\un x}_{1'},
y_1)$ we have to write down an evolution equation for the $S$-matrix
of the evolution of quadrupole $2,2',1,1'$. An evolution equation
involving a color quadrupole was derived before in \cite{CM} to
reproduce the BJKP equation \cite{BJKP} for four reggeons in the
framework of the dipole model \cite{dip}. The equation derived in
\cite{CM} corresponds to off-forward evolution for dipoles in the 
presence of a single quadrupole, with all the evolution included in
the dipoles. Therefore it should not be compared to the equation we
are about to write down, since we are interested in the evolution
inside the quadrupole.

\begin{figure}
\begin{center}
\epsfxsize=10cm
\leavevmode
\hbox{\epsffile{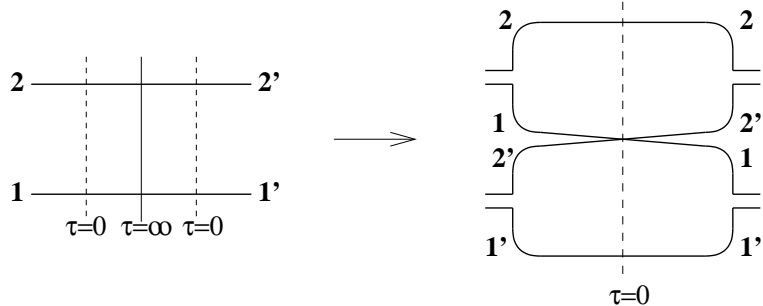}}
\end{center}
\caption{Redrawing the quadrupole interaction amplitude in the form 
convenient for including quantum evolution (see text).}
\label{qtr}
\end{figure}

We are going to derive an evolution equation for the quadrupole
$S$-matrix $Q({\un x}_2,{\un x}_{2'},{\un x}_1,{\un x}_{1'},y_1)$
including all the non-linear evolution effects. The initial condition
for evolution of $Q({\un x}_2,{\un x}_{2'},{\un x}_1,{\un
x}_{1'},y_1)$ is given by \eq{Q0}. To write down one step of the
evolution we first redraw the quadrupole as shown in
\fig{qtr}. Instead of the amplitude squared pictured on the left of
\fig{qtr} we will use a form of quadrupole cross section similar to
the forward amplitude shown on the right of \fig{qtr}. Obviously the
diagram on the right of \fig{qtr} preserves the color structure of the
quadrupole. All the interactions with the target in the graph of the
left happen along the dashed lines at time $\tau = 0$. On the right of
\fig{qtr} we merge two dashed lines from the graph on the left into
one dashed line with interactions in it. This way, a real interaction
with a nucleon where a single gluon is exchanged in each of the dashed
lines in the left graph of \fig{qtr} becomes a two-gluon exchange
(diffractive) interaction in the dashed line in the graph on the right
of \fig{qtr}. Again the picture is similar to the forward amplitude
calculation.

\begin{figure}
\begin{center}
\epsfxsize=12cm
\leavevmode
\hbox{\epsffile{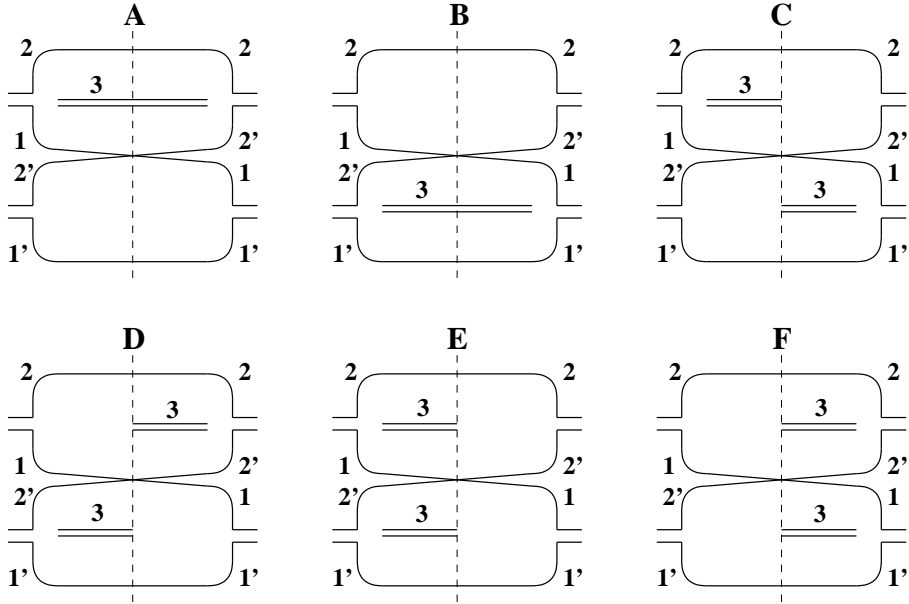}}
\end{center}
\caption{One step of the quadrupole evolution. }
\label{qev}
\end{figure}

One step of the quadrupole evolution in the representation of
\fig{qtr} is shown in \fig{qev}. The step consists of an emission of a single 
gluon $\# 3$. For instance, in \fig{qev}A the gluon $\# 3$ splits the
original quadrupole $2,2',1,1'$ into a quadrupole $3,2',1,1'$ and a
dipole $23$. \fig{qev}B gives a similar contribution. In
Figs. \ref{qev}C-F we drew the gluon $\# 3$ line as disconnected at
dashed line: the gluon line is indeed implied to be connected and
continuous. The disconnected line is just a notation which we borrowed
from \cite{CM} for, say, a gluon emitted in dipole $21$ and absorbed
in dipole $11'$ in \fig{qev}C. However, when connecting the two parts
of the gluon $\# 3$ line in Figs. \ref{qev}C-F one has to be careful
to pick up the leading large-$N_c$ contribution. For instance, in
\fig{qev}C the leading term consists of gluon $\# 3$ splitting the
original quadrupole $2,2',1,1'$ into a dipole $31$ and a quadrupole
$2,3,1',2'$. The dominant contribution of \fig{qev}E consists of gluon
$\# 3$ splitting the quadrupole $2,2',1,1'$ into "dipoles" $2,2',3$
and $1,1',3$, in which interactions with line $3$ cancel due to
real-virtual cancellations. Leading terms for other graphs in
\fig{qev} can be obtained in similar ways. Combining them with virtual
corrections yields
\ben
Q ({\un x}_2, {\un x}_{2'}, {\un x}_1, {\un x}_{1'}, y_1) \, = \, e^{-
\bas \, \ln \left( \frac{x_{21} x_{2'1'} x_{22'} x_{11'}}{\rho^4} \right) \, 
y_1} \, Q_0 ({\un x}_2, {\un x}_{2'}, {\un x}_1, {\un x}_{1'}) +
\een
\ben
+ \frac{\bas}{2 \, \pi} \, \int_0^{y_1} \, dy \, e^{- \bas \, 
\ln \left( \frac{x_{21} x_{2'1'} x_{22'} x_{11'}}{\rho^4} \right) \, 
(y_1 - y)} \, \int d^2 x_3 \, \bigg\{
\left( \frac{{\un x}_{32}}{x_{32}^2} - 
\frac{{\un x}_{31}}{x_{31}^2} \right) \cdot 
\left( \frac{{\un x}_{32}}{x_{32}^2} - 
\frac{{\un x}_{32'}}{x_{32'}^2} \right) \, S ({\un x}_2, {\un x}_3, y)
\een
\ben
\times \, Q ({\un x}_3, {\un x}_{2'}, {\un x}_1, {\un x}_{1'}, y) +
\left( \frac{{\un x}_{32'}}{x_{32'}^2} - 
\frac{{\un x}_{31'}}{x_{31'}^2} \right) \cdot 
\left( \frac{{\un x}_{31}}{x_{31}^2} - 
\frac{{\un x}_{31'}}{x_{31'}^2} \right) \, S ({\un x}_3, {\un x}_{1'}, y) 
\, Q ({\un x}_2, {\un x}_{2'}, {\un x}_1, {\un x}_{3}, y) -
\een
\ben
- \left( \frac{{\un x}_{32}}{x_{32}^2} - 
\frac{{\un x}_{31}}{x_{31}^2} \right) \cdot 
\left( \frac{{\un x}_{31}}{x_{31}^2} - 
\frac{{\un x}_{31'}}{x_{31'}^2} \right) \, S ({\un x}_3, {\un x}_{1}, y) 
\, Q ({\un x}_2, {\un x}_{2'}, {\un x}_3, {\un x}_{1'}, y) - 
\left( \frac{{\un x}_{32}}{x_{32}^2} - 
\frac{{\un x}_{32'}}{x_{32'}^2} \right) \cdot 
\left( \frac{{\un x}_{32'}}{x_{32'}^2} - 
\frac{{\un x}_{31'}}{x_{31'}^2} \right)
\een
\ben
\times \, S ({\un x}_3, {\un x}_{2'}, y) 
\, Q ({\un x}_2, {\un x}_{3}, {\un x}_1, {\un x}_{1'}, y) +
\left( \frac{{\un x}_{32}}{x_{32}^2} - 
\frac{{\un x}_{31}}{x_{31}^2} \right) \cdot 
\left( \frac{{\un x}_{32'}}{x_{32'}^2} - 
\frac{{\un x}_{31'}}{x_{31'}^2} \right) \, S ({\un x}_2, {\un x}_{2'}, y) 
\, S ({\un x}_1, {\un x}_{1'}, y)
\een
\be\label{Qev}
+ \left( \frac{{\un x}_{32}}{x_{32}^2} - 
\frac{{\un x}_{32'}}{x_{32'}^2} \right) \cdot 
\left( \frac{{\un x}_{31}}{x_{31}^2} - 
\frac{{\un x}_{31'}}{x_{31'}^2} \right) \, S ({\un x}_2, {\un x}_1, y) 
\, S ({\un x}_{2'}, {\un x}_{1'}, y) \bigg\}.
\ee
Note that due to real-virtual cancellations the last two terms in
\eq{Qev} corresponding to diagrams E and F in \fig{qev} contain only 
the dipole $S$-matrices in them. The relative signs of various terms
in \eq{Qev} are easier to determine in the representation of the
interaction on the left hand side of \fig{qtr} keeping in mind that
gluon emissions at $\tau < 0$ and $\tau > 0$ come in with different
signs.

As we have already mentioned the linearized version of \eq{Qev} (see
\eq{Qevlin} below) should be combined with Eq. (49) in \cite{CM} to complete
the description of the BJKP evolution in the framework of the dipole
model. (By linearizing \eq{Qev} we mean substituting $S=1-N$ in it
(see \eq{S}) while keeping only terms linear in $Q$ and in $N$, i.e.,
neglecting products like $Q N$ and $N N$.) However, as was discussed
in \cite{CM}, the contribution of linearized \eq{Qev} to BJKP
evolution is likely to be small: as we will see in the next
Subsection, the linearized version of \eq{Qev} is almost equivalent to
the BFKL equation. Therefore, its solution is likely to grow with
energy just like a single BFKL pomeron, which is much slower than the
double BFKL pomeron exchange. On the other hand, a solution of the
full BJKP evolution equation for four reggeons presented in \cite{KKM}
has an intercept much smaller than that of a single BFKL pomeron
exchange making the contribution of \eq{Qev} potentially important for
BJKP evolution in the dipole model.

Let us check \eq{Qev} for consistency with our earlier results. First
note that if ${\un x}_2 = {\un x}_{2'}$ the interactions with line
2/2' would cancel and the following equality should be true
\be\label{test1}
Q ({\un x}_2, {\un x}_{2}, {\un x}_1, {\un x}_{1'}, y_1) \, = \, S
({\un x}_1, {\un x}_{1'}, y_1) \, = \, 1 - N ({\un x}_1, {\un x}_{1'},
y_1).
\ee
>From Eqs. (\ref{test0}) and (\ref{N0}) we can see that \eq{test1} is
certainly true for the initial conditions for \eq{Qev} given by $Q_0$
from \eq{Q0}. Now, as one can explicitly check, putting ${\un x}_2 =
{\un x}_{2'}$ in \eq{Qev} (with $x_{22'} \rightarrow \rho$ in the
exponent) and assuming that \eq{test1} is true one readily recovers
\eq{eqN}. Thus \eq{Qev} consistently maps onto \eq{eqN} in the limit 
of \eq{test1}. 

Using \eq{test1} in \eq{d} and remembering that due to \eq{ngn} the
$S$-matrix of a gluon dipole $S_G$ can be expressed in terms of the
$S$-matrix of the quark dipole as
\be\label{SG}
S_G ({\un x}_0, {\un x}_1, y) \, = \, S^2 ({\un x}_0, {\un x}_1, y)
\ee
we observe that
\be\label{test2}
d ({\un x}_0, {\un x}_{0}, {\un x}_1 , {\un k}, y) \, = \, s 
({\un x}_0, {\un x}_1 , {\un k}, y),
\ee
which verifies that \eq{d} is consistent with \eq{sdef}.

\eq{Qev}, when solved to find $Q$, can be used to construct $d$ in \eq{d}, 
which can then be used as initial condition to the evolution equation
(\ref{Mev}). The quantity $M ({\un x}_2, {\un x}_{2'}, {\un x}_{\tilde
0}, Y; {\un k}_1, y_1)$ obtained this way can be used in \eq{2Gcl}
instead of $M_0$ along with $S$ from \eq{S} to obtain the evolved
contribution of case B considered above to the two-gluon production
cross section. Together with the contribution of the case A from
\eq{Apiece} it gives us the following expression for the double gluon 
production cross section in a quark dipole--nucleus scattering [with
${\un B} = ({\un x}_0 + {\un x}_{\tilde 0})/2$]
\ben
\frac{d \sigma^{q {\bar q} \, A \ \rightarrow \ q {\bar q} \, G_1 G_2 X}}
{d^2 k_1 \, dy_1 \, d^2 k_2 \, dy_2} ({\un x}_{0{\tilde 0}})
\bigg|_{y_2 \gg y_1}\, = \,
 \int \ d^2 B \ \bigg\{ n_2 ({\un x}_0, {\un x}_{\tilde 0}, Y; {\un
 x}_1 , {\un x}_{\tilde 1}, y_1, {\un x}_2, {\un x}_{\tilde 2}, y_2)
\een
\ben
\times \, d^2 x_1 \, d^2 x_{\tilde 1} \, d^2 x_2 \, d^2 x_{\tilde 2} \, s
 ({\un x}_1, {\un x}_{\tilde 1}, {\un k}_1, y_1) \, s ({\un x}_{2},
 {\un x}_{\tilde 2}, {\un k}_2, y_2)
+ n_1 ({\un x}_0, {\un x}_{\tilde 0}, Y; {\un x}_1 , {\un
x}_{\tilde 1}, y_2) \, d^2 x_1 \, d^2 x_{\tilde 1} 
\een
\ben
\times \, \frac{\bas}{(2 \pi)^3}
\, \int \, d^2 x_2 \, d^2 x_{2'} \, 
e^{- i {\underline k}_2 \cdot {\underline x}_{22'}} \,
\bigg[ \left( \frac{{\un x}_{21}}{x_{21}^2} - 
\frac{{\un x}_{2{\tilde 1}}}{x_{2{\tilde 1}}^2} \right) \cdot 
\left( \frac{{\un x}_{2'1}}{x_{2'1}^2} - 
\frac{{\un x}_{2'{\tilde 1}}}{x_{2'{\tilde 1}}^2} \right) \, 
M ({\un x}_2, {\un x}_{2'}, {\un x}_{\tilde 1}, y_2; {\un k}_1,
y_1) \, S({\un x}_2, {\un x}_{2'}, y_2) 
\een
\ben
- \left( \frac{{\un x}_{21}}{x_{21}^2} - 
\frac{{\un x}_{2{\tilde 1}}}{x_{2{\tilde 1}}^2} \right) \cdot 
\frac{{\un x}_{2'1}}{x_{2'1}^2} \, 
M ({\un x}_2, {\un x}_1, {\un x}_{\tilde 1}, y_2; {\un k}_1,
y_1) \, S({\un x}_2, {\un x}_1, y_2)
- \left( \frac{{\un x}_{2'1}}{x_{2'1}^2} - 
\frac{{\un x}_{2'{\tilde 1}}}{x_{2'{\tilde 1}}^2} \right) \cdot 
\frac{{\un x}_{21}}{x_{21}^2} 
\een
\be\label{2Gincl}
\times \, M ({\un x}_1, {\un x}_{2'}, {\un x}_{\tilde 1}, y_2; {\un k}_1,
y_1) \, S({\un x}_1, {\un x}_{2'}, y_2) 
+ \frac{{\un x}_{21}}{x_{21}^2} \cdot \frac{{\un x}_{2'1}}{x_{2'1}^2}
\, M ({\un x}_1, {\un x}_1, {\un x}_{\tilde 1}, y_2; {\un k}_1, y_1) \,
+ \, (1 \leftrightarrow {\tilde 1}) \bigg] \bigg\}.
\ee
\eq{2Gincl} is the central result of this Section. Together with Eqs. 
(\ref{disincl}), (\ref{Qev}), (\ref{d}), (\ref{Mev}), (\ref{n1}),
(\ref{n2}) and (\ref{eqN}) it gives us the expression for the
two-gluon inclusive cross section for DIS on a nucleus with the
effects of nonlinear evolution (\ref{eqN}) included.

\begin{figure}
\begin{center}
\epsfxsize=16.5cm
\leavevmode
\hbox{\epsffile{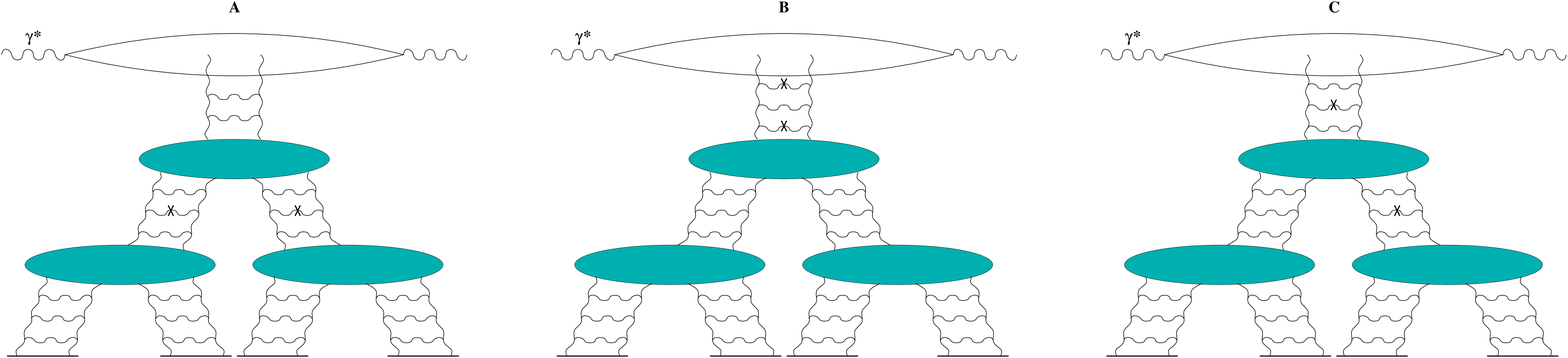}}
\end{center}
\caption{Feynman diagrams corresponding to double gluon production 
cross section given by \eq{2Gincl}. Emitted gluons are denoted by
crosses.  }
\label{fan2}
\end{figure}

The structure of \eq{2Gincl} is illustrated in \fig{fan2}, where, if
one pictures the evolution of \eq{eqN} as resumming fan diagrams, the
diagrams correspond to the first (case A) and the second (cases B
and C) terms in \eq{2Gincl}. The first term in \eq{2Gincl}
corresponds to splitting of the original linear evolution in two,
which is described by \eq{n2} for $n_2$. Then each of the two ladders
independently produces a gluon with all the possible splittings
happening afterwards. This is illustrated in \fig{fan2}A.  The
second term in \eq{2Gincl} corresponds to nonlinear evolution
successively producing both gluons, after which all possible
splittings are allowed, as shown in Figs. \ref{fan2}B and \ref{fan2}C,
where we have divided the nonlinear evolution into the linear
(Fig. \ref{fan2}B) and nonlinear (Fig. \ref{fan2}C) parts. The linear
evolution in this second term in \eq{2Gincl} is given by $n_1$ from
\eq{n1} and by $M$ from the linear part of \eq{Mev}. This linear
evolution leads to production of both gluons $\# 2$ and $\# 1$ and is
illustrated in \fig{fan2}B. The initial conditions for \eq{Mev} are
nonlinear, given by Eqs. (\ref{d}) and (\ref{Qev}). They include
ladder splittings and are pictured by the fan diagram in the lower
part of \fig{fan2}B. One should note, however, that \eq{Mev}, while
being linear in $M$, has extra factors of $1-N$ on its right hand
side. That means that evolution of $M$ includes ladder splittings
between the gluons $\# 2$ and $\# 1$, one of which is shown in
\fig{fan2}C. There the evolution leading to creation of gluon $\# 2$ 
is still linear since it is still given by $n_1$ in the second term on
the right hand side of \eq{2Gincl}. However, since the evolution in
the rapidity interval between the emitted gluons (evolution of $M$) is
nonlinear, splittings are allowed between the gluons $\# 2$ and $\#
1$, as depicted in \fig{fan2}C.

The diagrams A and B in \fig{fan2} are the same as would have been
expected from AGK cutting rules \cite{agk} (see also \cite{kt} and
\cite{KL} for similar correspondence between the dipole model results
and AGK rules expectations). However, the diagram C in \fig{fan2},
while being included in \eq{2Gincl}, is prohibited by AGK cutting
rules. Therefore, we seem to observe a direct violation of the AGK
rules in QCD. Since AGK rules have never been proven for
QCD, one should not be too surprised that they do not work here. It is
interesting to note that AGK violation sets in at the level of the
2-gluon production: single gluon inclusive production cross section
calculated in \cite{kt} adheres to AGK rules and so does the
diffractive DIS cross section calculated in \cite{KL}.

The violation of AGK cutting rules in \eq{2Gincl} is due to non-linear
terms in \eq{Mev}, which are in turn due to late time (after the
interaction) gluon emissions at light cone times $\tau > 0$. These
terms were not important for the calculation of the total cross
section in the dipole model \cite{dip}: there they were found to
cancel \cite{CM}. Thus if one would try to construct an analogy
between the fan diagrams \cite{glr} and dipole calculations \cite{bk}
based on the correspondence of total cross sections, one would omit
such terms. Since the fan diagrams seem to adhere to AGK rules, this
omission would lead to the erroneous conclusion that AGK rules should
work for the production cross section. However, as we have seen above,
these late time emissions are important for single \cite{kt} and
double inclusive gluon production, violating the AGK rules for the
latter. What appears to fail here is the one-to-one correspondence
between the fan diagrams and dipole calculations.

Another difference between our result (\ref{2Gincl}) and the direct
application of AGK rules to calculating inclusive cross section done
in \cite{braun} is that nonlinear splittings may start exactly at the
point in rapidity when the softer of the produced gluons is emitted in
the diagram B or exactly at the point of emission of both gluons $\#
1$ and $\# 2$ in the diagram A. Similar discrepancy was already
observed when comparing the single gluon inclusive cross section
calculated in \cite{kt} to the results of \cite{braun}.  

In comparing our result with the formula obtained in \cite{braun} (see
Eqs. (30) and (36) there) we note that in a general case we could not
cast \eq{2Gincl} in the $k_T$-factorized form of \cite{braun}. Again,
this distinguishes the case of two-gluon production considered here
from the case of single gluon production from \cite{kt}.

Indeed our result, given by \eq{2Gincl}, is rather complicated,
especially keeping in mind that one has to first solve evolution
equations (\ref{n1}), (\ref{n2}), (\ref{eqN}), (\ref{Qev}), and
(\ref{Mev}) in order to obtain the desired two-particle production
cross section. In order to make \eq{2Gincl} easier to implement, it is
highly desirable to find some way of simplifying it. Unfortunately, we
could not find any simplification of \eq{2Gincl} in the general
case. Nevertheless, in certain kinematic regimes \eq{2Gincl} may be
simplified. For instance, if the center of mass energy of the
collision is not too high or if the transverse momenta of the produced
gluons are sufficiently large ($|{\un k}_1|, |{\un k}_2| \gsim Q_s$),
the nonlinear saturation effects, such as ladder splittings, could be
neglected. This implies that the diagrams in Figs. \ref{fan2} A and C
are small with the linear part of the diagram B dominating the cross
section. This is the well-known leading twist result
\cite{lr}, which we will derive from our \eq{2Gincl} in Sect. IIIC below. 

In the opposite kinematic regime of very large center of mass energy
of the collision and not too high gluons' transverse momenta
saturation effects become important. There one can note that in
Figs. \ref{fan2} B and C the evolution between the projectile and the
(harder) gluon $\# 2$ is linear and is given by a single BFKL ladder
exchange. On the other hand one can show \cite{braun} that the diagram
in \fig{fan2}A is dominated by the contribution where the triple
pomeron vertex in the evolution between the projectile and the gluon
$\# 2$ is all the way up at the projectile's rapidity. Therefore, the
evolution between the projectile and the gluon $\# 2$ in \fig{fan2}A
is given by a double pomeron exchange, and is thus energetically more
favorable than the single pomeron exchange of Figs. \ref{fan2} B and
C. Since the rest of the three diagrams is parametrically the same,
one concludes that \fig{fan2}A dominates in this regime, as was
originally shown in \cite{braun}. Keeping only the corresponding first
term on the right hand side of \eq{2Gincl} would significantly
simplify the calculation of the cross section: since an analytical
solution of \eq{n2} for $n_2$ exists \cite{dip}, one would only need
to find a solution of \eq{eqN}, for which there is a number of
analytical and numerical results in the literature. However, one has
to be careful in neglecting the diagrams in Figs. \ref{fan2} B and
C. If one is interested in azimuthal two-particle correlations of the
produced gluons, than the contributions of graphs in Figs. \ref{fan2}
B and C might be more important than the contribution of \fig{fan2}A
even deep inside the saturation region \cite{klm2}. 

\subsection{Recovering the Leading Twist Result}

Let us show that \eq{2Gincl} reduces to the usual ``leading-twist''
$k_T$-factorization result \cite{lr} in the limit of large transverse
momenta of the produced gluons. Large transverse momenta correspond to
small transverse distances. For short transverse distances all the
evolution equations written above should be linearized since all the
non-linearities would be negligibly small. Therefore we can right away
neglect the first term on the right hand side of \eq{2Gincl}, which
contains a splitting (which is a non-linearity) in the evolution
between the target and emitted gluons as shown in \fig{fan2}A. In the
remaining second term on the right of \eq{2Gincl} we can put all $S
\approx 1$ to obtain the linearized expression
\ben
\frac{d \sigma^{q {\bar q} \, A \ \rightarrow \ q {\bar q} \, G_1 G_2 X}}
{d^2 k_1 \, dy_1 \, d^2 k_2 \, dy_2} ({\un x}_{0{\tilde 0}})
\bigg|_{LO} \, \approx \, \int \ d^2 B \
n_1 ({\un x}_0, {\un x}_{\tilde 0}, Y; {\un x}_1 , {\un
x}_{\tilde 1}, y_2) \, d^2 x_1 \, d^2 x_{\tilde 1} 
\een
\ben
\times \, \frac{\bas}{(2 \pi)^3}
\, \int \, d^2 x_2 \, d^2 x_{2'} \, 
e^{- i {\underline k}_2 \cdot {\underline x}_{22'}} \,
\bigg[ \left( \frac{{\un x}_{21}}{x_{21}^2} - 
\frac{{\un x}_{2{\tilde 1}}}{x_{2{\tilde 1}}^2} \right) \cdot 
\left( \frac{{\un x}_{2'1}}{x_{2'1}^2} - 
\frac{{\un x}_{2'{\tilde 1}}}{x_{2'{\tilde 1}}^2} \right) \, 
M ({\un x}_2, {\un x}_{2'}, {\un x}_{\tilde 1}, y_2; {\un k}_1,
y_1) \bigg|_{lin}
\een
\ben
- \left( \frac{{\un x}_{21}}{x_{21}^2} - 
\frac{{\un x}_{2{\tilde 1}}}{x_{2{\tilde 1}}^2} \right) \cdot 
\frac{{\un x}_{2'1}}{x_{2'1}^2} \, 
M ({\un x}_2, {\un x}_1, {\un x}_{\tilde 1}, y_2; {\un k}_1,
y_1) \bigg|_{lin} \,
- \left( \frac{{\un x}_{2'1}}{x_{2'1}^2} - 
\frac{{\un x}_{2'{\tilde 1}}}{x_{2'{\tilde 1}}^2} \right) \cdot 
\frac{{\un x}_{21}}{x_{21}^2} \, 
 M ({\un x}_1, {\un x}_{2'}, {\un x}_{\tilde 1}, y_2; {\un k}_1, y_1)
 \bigg|_{lin}
\een
\be\label{2Glin1}
+ \frac{{\un x}_{21}}{x_{21}^2} \cdot \frac{{\un x}_{2'1}}{x_{2'1}^2}
\, M ({\un x}_1, {\un x}_1, {\un x}_{\tilde 1}, y_2; {\un k}_1, y_1) 
\bigg|_{lin}
\, + \, (1 \leftrightarrow {\tilde 1}) \bigg].
\ee
To linearize the evolution equation for $M$ (\eq{Mev}) we start by
linearizing its initial condition given by $d$. It is determined by
\eq{d} with $Q$ given by \eq{Qev}. Therefore, we start with the 
initial conditions for \eq{Qev} given by \eq{Q0}. Expanding \eq{Q0} to
the lowest order in the transverse separations (or, equivalently, in
$Q_{s0}$) we obtain
\be\label{Q0n0}
Q_0 ({\un x}_2, {\un x}_{2'}, {\un x}_1, {\un x}_{1'}) \bigg|_{LO} \,
\approx \, 1 - n_0 ({\un x}_{21}) - n_0 ({\un x}_{2'1'}) 
- n_0 ({\un x}_{22'}) - n_0 ({\un x}_{11'}) + n_0 ({\un x}_{21'}) +
n_0 ({\un x}_{2'1}),
\ee
where the two-gluon exchange amplitude $n_0$ is given by the first term 
in the expansion of $N_0$ from \eq{N0}
\be\label{n0}
n_0 ({\un x}_{21}) \, = \, \frac{1}{4} \, x_{21}^2 \, Q_{s0}^2 \, 
\ln \frac{1}{x_{21} \, \Lambda}.
\ee
Let us assume that \eq{Qev} independently includes linear BFKL
evolution in each of the $n_0$'s in \eq{Q0n0}, such that the fully
evolved linearized quadrupole amplitude $Q$ is given by
\ben
q ({\un x}_2, {\un x}_{2'}, {\un x}_1, {\un x}_{1'},Y)  \, \equiv \, 
Q ({\un x}_2, {\un x}_{2'}, {\un x}_1, {\un x}_{1'},Y)  \bigg|_{lin} \,
= \, 1 - n ({\un x}_{2}, {\un x}_1, Y) - n ({\un x}_{2'}, {\un x}_{1'},Y) 
- n ({\un x}_{2}, {\un x}_{2'}, Y)
\een
\be\label{Qans}
 - n ({\un x}_{1}, {\un x}_{1'}, Y)
+ n ({\un x}_{2}, {\un x}_{1'},Y) + n ({\un x}_{2'}, {\un x}_{1}, Y)
\ee
with $n$ determined by the linearized version of \eq{eqN}
corresponding to BFKL evolution \cite{bfkl}
\ben
  n({\underline x}_{0}, {\un x}_{\tilde 0}, Y) = n_0 ({\underline
  x}_{0{\tilde 0}}) \, e^{- 2 \, \bas \, \ln
  \left( \frac{x_{0{\tilde 0}}}{\rho} \right) Y} + \frac{\bas}{2 \, \pi}
  \int_0^Y d y \ e^{ - 2\, \bas \, \ln \left(
  \frac{x_{0{\tilde 0}}}{\rho} \right) (Y - y)}
\een
\be\label{bfkl}
\times \int_\rho d^2 x_2 \frac{x_{0{\tilde 0}}^2}{x_{20}^2 
x_{2{\tilde 0}}^2} \, [ n ({\underline x}_{0},{\un x}_2, y) + n
({\underline x}_{2},{\un x}_{\tilde 0}, y)].
\ee
After lengthy algebra one can show that $q({\un x}_2, {\un x}_{2'},
{\un x}_1, {\un x}_{1'},Y)$ from \eq{Qans} satisfies the linearized
version of \eq{Qev}
\ben
q ({\un x}_2, {\un x}_{2'}, {\un x}_1, {\un x}_{1'}, y_1) \, = \, e^{-
\bas \, \ln \left( \frac{x_{21} x_{2'1'} x_{22'} x_{11'}}{\rho^4} \right) \, 
y_1} \, Q_0 ({\un x}_2, {\un x}_{2'}, {\un x}_1, {\un x}_{1'})
\bigg|_{LO} +
\een
\ben
+ \frac{\bas}{2 \, \pi} \, \int_0^{y_1} \, dy \, e^{- \bas \, 
\ln \left( \frac{x_{21} x_{2'1'} x_{22'} x_{11'}}{\rho^4} \right) \, 
(y_1 - y)} \, \int d^2 x_3 \, \bigg\{
\left( \frac{{\un x}_{32}}{x_{32}^2} - 
\frac{{\un x}_{31}}{x_{31}^2} \right) \cdot 
\left( \frac{{\un x}_{32}}{x_{32}^2} - 
\frac{{\un x}_{32'}}{x_{32'}^2} \right) 
\een
\ben
\times \, q ({\un x}_3, {\un x}_{2'}, {\un x}_1, {\un x}_{1'}, y) +
\left( \frac{{\un x}_{32'}}{x_{32'}^2} - 
\frac{{\un x}_{31'}}{x_{31'}^2} \right) \cdot 
\left( \frac{{\un x}_{31}}{x_{31}^2} - 
\frac{{\un x}_{31'}}{x_{31'}^2} \right) 
\, q ({\un x}_2, {\un x}_{2'}, {\un x}_1, {\un x}_{3}, y) -
\een
\ben
- \left( \frac{{\un x}_{32}}{x_{32}^2} - 
\frac{{\un x}_{31}}{x_{31}^2} \right) \cdot 
\left( \frac{{\un x}_{31}}{x_{31}^2} - 
\frac{{\un x}_{31'}}{x_{31'}^2} \right) 
\, q ({\un x}_2, {\un x}_{2'}, {\un x}_3, {\un x}_{1'}, y) - 
\left( \frac{{\un x}_{32}}{x_{32}^2} - 
\frac{{\un x}_{32'}}{x_{32'}^2} \right) \cdot 
\left( \frac{{\un x}_{32'}}{x_{32'}^2} - 
\frac{{\un x}_{31'}}{x_{31'}^2} \right)
\een
\ben
\times \, q ({\un x}_2, {\un x}_{3}, {\un x}_1, {\un x}_{1'}, y) +
\left( \frac{{\un x}_{32}}{x_{32}^2} - 
\frac{{\un x}_{31}}{x_{31}^2} \right) \cdot 
\left( \frac{{\un x}_{32'}}{x_{32'}^2} - 
\frac{{\un x}_{31'}}{x_{31'}^2} \right) \, [1 - n ({\un x}_2, {\un x}_{2'}, y) 
- n ({\un x}_1, {\un x}_{1'}, y)]
\een
\be\label{Qevlin}
+ \left( \frac{{\un x}_{32}}{x_{32}^2} - 
\frac{{\un x}_{32'}}{x_{32'}^2} \right) \cdot 
\left( \frac{{\un x}_{31}}{x_{31}^2} - 
\frac{{\un x}_{31'}}{x_{31'}^2} \right) \, [1 - n ({\un x}_2, {\un x}_1, y) 
- n ({\un x}_{2'}, {\un x}_{1'}, y) ] \bigg\}.
\ee
This proves that the ansatz of \eq{Qans} is indeed the correct
linearized quadrupole amplitude $Q$. Therefore, to construct the
initial conditions for the linearized version of \eq{Mev} we should
substitute $q({\un x}_2, {\un x}_{2'}, {\un x}_1, {\un x}_{1'},Y)$
from \eq{Qans} into the linearized version of \eq{d} obtaining
\ben
d_{lin} ({\un x}_2, {\un x}_{2'}, {\un x}_{\tilde 0} , {\un k}_1, y_1) \, = \,
\frac{2 \, \bas}{(2 \pi)^3} \, \int \, d^2 x_1 \, d^2
x_{1'} \, e^{- i {\un k}_1 \cdot {\un x}_{11'}} \, \bigg\{ 
\frac{{\un x}_{12}}{x_{12}^2} \cdot 
\frac{{\un x}_{1'2'}}{x_{1'2'}^2} \ \bigg[ 
n ( {\un x}_{2'}, {\un x}_1, y_1) + n ({\un x}_2, {\un x}_{1'}, y_1) 
\een
\ben
- n ({\un x}_2, {\un x}_{2'}, y_1) - n ({\un x}_1, {\un x}_{1'}, y_1) \, \bigg]
+ \frac{{\un x}_{1{\tilde 0}}}{x_{1{\tilde 0}}^2} \cdot 
\frac{{\un x}_{1'{\tilde 0}}}{x_{1'{\tilde 0}}^2} 
\bigg[ n ({\un x}_1, {\un x}_{\tilde 0}, y_1)  
+ n ({\un x}_{1'}, {\un x}_{\tilde 0}, y_1) - n ({\un x}_1, {\un
x}_{1'}, y_1) \bigg]
\een
\ben
 - \frac{{\un x}_{12}}{x_{12}^2} \cdot 
\frac{{\un x}_{1'{\tilde 0}}}{x_{1'{\tilde 0}}^2} \, 
\bigg[ n ({\un x}_2, {\un x}_{1'},  y_1) + 
n ({\un x}_1, {\un x}_{\tilde 0}, y_1) - n ({\un x}_2, {\un x}_{\tilde
0}, y_1) - n ({\un x}_{1}, {\un x}_{1'}, y_1) \bigg]
\een
\be\label{dlin1}
- \frac{{\un x}_{1{\tilde 0}}}{x_{1{\tilde 0}}^2} \cdot \frac{{\un
x}_{1'2'}}{x_{1'2'}^2} \, \bigg[ n ({\un x}_1, {\un x}_{2'}, y_1) + n
({\un x}_{1'}, {\un x}_{\tilde 0}, y_1) - n ({\un x}_{2'}, {\un
x}_{\tilde 0}, y_1) - n ({\un x}_{1}, {\un x}_{1'}, y_1) \bigg]
\bigg\}.
\ee
Here we consider scattering on a large nucleus. The amplitude $n ({\un
x}_1, {\un x}_{1'}, y)$ depends on the transverse size of the
quark-antiquark pair $x_{11'}$ as well as on the transverse position
of the dipole. However, the transverse position of the dipole is given
by the overall impact parameter ${\un B}$ in the scattering
process. The integrals over ${\un x}_1$ and ${\un x}_{1'}$ in
\eq{dlin1}, while formally going out to infinity in the transverse 
direction, are indeed effectively limited to the typical hadronic size
on which the concept of a gluon still makes sense. The dependence of
$n$ on ${\un B}$ is smooth for a large nucleus, slowly varying on
transverse distances of the order of the typical hadronic
size. Therefore we write (see \cite{kt} for a similar approximation)
\be\label{cyl}
n ({\un x}_1, {\un x}_2, y) \, \approx \, n ({\un x}_{12}, {\un B},
y).
\ee
With the help of \eq{cyl} we rewrite \eq{dlin1} as
\be\label{dlin2}
d_{lin} ({\un x}_2, {\un x}_{2'}, {\un x}_{\tilde 0} , {\un k}_1, y_1)
\, = \, 
\frac{\bas}{2 \pi^2} \, \frac{1}{{\un k}_1^2} \, \int \, d^2 z \, 
n({\un z}, {\un B}, y_1) \,
\nabla^2_z \, \left( e^{- i {\un k}_1 \cdot {\un z}} \, \ln 
\frac{|{\un z} - {\un x}_{2{\tilde 0}}| \, |{\un z} + {\un x}_{2'{\tilde 0}}|}
{|{\un z} - {\un x}_{22'}| \, |{\un z}|} \right),
\ee
where $\nabla^2_z$ is the transverse coordinate gradient squared.
\eq{dlin2} is the initial condition for the linearized version of 
\eq{Mev}. The latter can be obtained from \eq{Mev} by putting all $N=0$ in it, 
which yields
\ben
m ({\un x}_2, {\un x}_{2'}, {\un x}_{\tilde 0}, Y; {\un k}_1, y_1) \,
= \, e^{- \bas \, \ln \left( \frac{x_{2{\tilde 0}} x_{2'{\tilde 0}}
x_{22'}}{\rho^3} \right) \, (Y-y_1)} \, d_{lin} ({\un x}_2, {\un
x}_{2'}, {\un x}_{\tilde 0}, {\un k}_1, y_1) +
\een
\ben
+ \frac{\bas}{2 \pi} \, \int d^2 x_4 \int_{y_1}^Y dy \, e^{- \bas \,
\ln \left( \frac{x_{2{\tilde 0}} x_{2'{\tilde 0}} x_{22'}}{\rho^3}
\right) \, (Y-y)} \, \bigg\{
\left( \frac{{\un x}_{42}}{x_{42}^2} - 
\frac{{\un x}_{4{\tilde 0}}}{x_{4{\tilde 0}}^2} \right) \cdot 
\left( \frac{{\un x}_{42'}}{x_{42'}^2} - 
\frac{{\un x}_{4{\tilde 0}}}{x_{4{\tilde 0}}^2} \right) 
\een
\ben
\times 
\bigg[ m ({\un x}_2, {\un x}_{2'}, {\un x}_4, y; {\un k}_1, y_1) 
+ \int d^2 x_a d^2 x_b \, n_1 ({\un x}_4, {\un x}_{\tilde 0}, y;
{\un x}_a, {\un x}_b, y_1) \, s ({\un x}_a, {\un x}_b, {\un k}_1, y_1)
 \bigg] - 
\een
\ben
- \left( \frac{{\un x}_{42}}{x_{42}^2} - 
\frac{{\un x}_{4{\tilde 0}}}{x_{4{\tilde 0}}^2} \right) \cdot 
\left( \frac{{\un x}_{42'}}{x_{42'}^2} - 
\frac{{\un x}_{42}}{x_{42}^2} \right) \, 
m ({\un x}_4, {\un x}_{2'}, {\un x}_{\tilde 0}, y; {\un k}_1, y_1) - 
\een
\be\label{mevl}
- \left( \frac{{\un x}_{42}}{x_{42}^2} - 
\frac{{\un x}_{42'}}{x_{42'}^2} \right) \cdot 
\left( \frac{{\un x}_{42'}}{x_{42'}^2} - 
\frac{{\un x}_{4{\tilde 0}}}{x_{4{\tilde 0}}^2} \right) \, 
m ({\un x}_2, {\un x}_4, {\un x}_{\tilde 0}, y; {\un k}_1, y_1) \bigg\} ,
\ee
where we have introduced a linearized amplitude $M$ denoted by
\be
m ({\un x}_2, {\un x}_{2'}, {\un x}_{\tilde 0}, Y; {\un k}_1, y_1) \,
\equiv \, M ({\un x}_2, {\un x}_{2'}, {\un x}_{\tilde 0}, Y; {\un
k}_1, y_1) \bigg|_{lin}.
\ee
The form of \eq{dlin2} provides us with the following ansatz for the
solution of \eq{mevl}:
\be\label{msol}
m ({\un x}_2, {\un x}_{2'}, {\un x}_{\tilde 0}, Y; {\un k}_1, y_1) \,
= \, f ({\un x}_{2{\tilde 0}}, Y; {\un k}_1, y_1) + f ({\un
x}_{2'{\tilde 0}}, Y; {\un k}_1, y_1) - f ({\un x}_{22'}, Y; {\un
k}_1, y_1)
\ee
with $f$ some unknown functions. Substituting the ansatz of \eq{msol}
into \eq{mevl} one can see that it is a solution of \eq{mevl} iff
\be
f ({\un x}_{21}, Y; {\un k}_1, y_1) \, = \, \frac{1}{2} \, \int d^2
x_a d^2 x_b \, n_1 ({\un x}_2, {\un x}_{1}, Y; {\un x}_a, {\un x}_b,
y_1) \, s ({\un x}_a, {\un x}_b, {\un k}_1, y_1).
\ee
The final answer for $m$ is
\ben
m ({\un x}_2, {\un x}_{2'}, {\un x}_{\tilde 0}, Y; {\un k}_1, y_1) \,
= \, \frac{1}{2} \, \int d^2 x_a d^2 x_b \, [n_1 ({\un x}_2, {\un
x}_{\tilde 0}, Y; {\un x}_a, {\un x}_b, y_1) + n_1 ({\un x}_{2'}, {\un
x}_{\tilde 0}, Y; {\un x}_a, {\un x}_b, y_1)
\een
\be\label{mM}
- n_1 ({\un x}_{2}, {\un x}_{2'}, Y;
{\un x}_a, {\un x}_b, y_1)] \, s ({\un x}_a, {\un x}_b, {\un k}_1, y_1).
\ee
Similar to \eq{cyl} we rewrite
\be
n_1 ({\un x}_{2}, {\un x}_{2'}, Y; {\un x}_a, {\un x}_b, y_1) \,
\longrightarrow \, n_1 ({\un x}_{22'}, {\un B}, Y; {\un x}_a, {\un
x}_b, y_1)
\ee
with the impact parameter ${\un B}$. Again we assume that for a large
nucleus $n_1$ is a slowly varying function of ${\un B}$. Using \eq{mM}
as $M$ in \eq{2Glin1} than yields
\ben
\frac{d \sigma^{q {\bar q} \, A \ \rightarrow \ q {\bar q} \, G_1 G_2 X}}
{d^2 k_1 \, dy_1 \, d^2 k_2 \, dy_2} ({\un x}_{0{\tilde 0}})
\bigg|_{LO} \, \approx \, \int \ d^2 B \
n_1 ({\un x}_0, {\un x}_{\tilde 0}, Y; {\un x}_1 , {\un
x}_{\tilde 1}, y_2) \, d^2 x_1 \, d^2 x_{\tilde 1} \, 
\frac{\bas}{2 \, (2 \pi)^2}
\een
\be\label{2Glin2}
\times  \, \frac{1}{{\un k}_2^2}
\, \int \, d^2 z \, 
e^{- i {\un k}_2 \cdot {\un z}} \, \ln \left( 
\frac{|{\un z} - {\un x}_{1{\tilde 1}}| \, |{\un z} + {\un x}_{1{\tilde 1}}|}
{|{\un z}|^2} \right) \, d^2 x_a d^2 x_b \, \nabla^2_z \, n_1 ({\un
z}, {\un B}, y_2; {\un x}_a, {\un x}_b, y_1) \, s ({\un
x}_a, {\un x}_b, {\un k}_1, y_1).
\ee
Similarly we rewrite \eq{sdef} as
\be\label{sfact}
s ({\un x}_a, {\un x}_b, {\un k}_1, y_1) \, = \, \frac{\bas}{(2 \pi)^2} 
\, \frac{1}{{\un k}_1^2} \, \int \, d^2 w \,  
e^{- i {\un k}_1 \cdot {\un w}} \, \ln  \left(
\frac{|{\un w} - {\un x}_{ab}| \, |{\un w} + {\un x}_{ab}|}
{|{\un w}|^2} \right)  \, \nabla^2_w \, N_G ({\un w}, {\un B}, y_1).
\ee
In the linear regime
\be\label{nN}
N_G ({\un w}, {\un B}, y_1) \, \approx \, 2 \, n ({\un w}, {\un B}, y_1)
\ee
with $n$ taken from \eq{bfkl}. Therefore, linearized $s$ can be
obtained from \eq{sfact} using \eq{nN} 
\be\label{slin}
s_{lin} ({\un x}_a, {\un x}_b, {\un k}_1, y_1) \, = \, \frac{\bas}{2 \pi^2} 
\, \frac{1}{{\un k}_1^2} \, \int \, d^2 w \,  
e^{- i {\un k}_1 \cdot {\un w}} \, \ln \left( 
\frac{|{\un w} - {\un x}_{ab}| \, |{\un w} + {\un x}_{ab}|}
{|{\un w}|^2} \right) \, \nabla^2_w \, n ({\un w}, {\un B}, y_1).
\ee
Defining ${\un b}_{ab} = ({\un x}_a + {\un x}_b)/2$ we relabel the
variables of $n_1$ in \eq{2Glin2} as \cite{dip,kt}
\be
n_1 ({\un z}, {\un B}, y_2; {\un x}_a, {\un x}_b, y_1) \, \longrightarrow \, 
n_1 ({\un z}, {\un x}_{ab}, {\un B}- {\un b}_{ab}, y_2 - y_1).
\ee
Than the integrals over ${\un x}_a$ and ${\un x}_b$ in \eq{2Glin2} can
be written as
\ben
\int \, d^2 x_{ab} \, d^2 b_{ab} \, \nabla^2_z \, 
 n_1 ({\un z}, {\un x}_{ab}, {\un B}- {\un b}_{ab}, y_2 - y_1) \,
 s_{lin} ({\un x}_a, {\un x}_b, {\un k}_1, y_1) 
\een
\be\label{term1}
 \, = \, \int \, d^2
 x_{ab} \, d^2 b_{ab} \, \nabla^2_z \, n_1 ({\un z}, {\un x}_{ab},
 {\un b}_{ab}, y_2 - y_1) \, s_{lin} ({\un x}_a, {\un x}_b,
 {\un k}_1, y_1),
\ee
where we have put the index indicating that we have to use a
linearized amplitude $s$ from \eq{slin} and shifted ${\un b}_{ab}$ by
${\un B}$.  Using the explicit solution of \eq{n1}
\cite{dip}
\be
\int \, d^2 b_{ab} \, n_1 ({\un z}, {\un x}_{ab}, 
{\un b}_{ab}, y_2 - y_1) \, = \,
\frac{1}{2 \pi \, {\un x}_{ab}^2} \, \int \frac{d
\lambda}{2 \pi i} \, e^{2 \, \bas \, \chi(\lambda) \, (y_2-y_1)} \, 
\left( \frac{z}{x_{ab}} \right)^\lambda
\ee
with the eigenvalue of the BFKL equation \cite{bfkl,dip}
\be
\chi (\lambda) \, = \, \psi (1) - \frac{1}{2} \, \psi (\lambda) - 
\frac{1}{2} \, \psi (1-\lambda)
\ee
and with the help of \eq{slin} we rewrite \eq{term1} as
\be\label{term2}
\frac{1}{2 \pi^2} \, \nabla^2_z \, \int \, d^2 w \, 
\left( \int \, d^2 b_{ab} \, 
\frac{1}{\nabla^4_w}   n_1 ({\un z}, {\un w}, {\un b}_{ab}, y_2 - y_1)
\right) \, {\hat L}_{k_1} ({\un w}) \, n ({\un w}, {\un B}, y_1),
\ee
where we have defined the operator for Lipatov's effective vertex
\cite{bfkl,braun}
\be
{\hat L}_{k} ({\un z}) \, \equiv \, \frac{4 \, \pi \, \bas}{{\un k}^2}
\, \stackrel{\textstyle\leftarrow}{\nabla}^2_z \, e^{- i {\underline k} \cdot
{\underline z}} \, \stackrel{\textstyle\rightarrow}{\nabla}^2_z .
\ee
Performing the integrations over ${\un x}_1$ and ${\un x}_{1'}$ in
\eq{2Glin2} in a similar manner we finally obtain
\ben
\frac{d \sigma^{q {\bar q} \, A \ \rightarrow \ q {\bar q} \, G_1 G_2 X}}
{d^2 k_1 \, dy_1 \, d^2 k_2 \, dy_2} ({\un x}_{0{\tilde 0}})
\bigg|_{LO} \, \approx \, \frac{1}{(2 \pi)^4} \, \int \ d^2 B \, d^2 z \, 
d^2 w \, \left( \int \, d^2 b_2 \, 
\frac{1}{\nabla^4_z}  n_1 ({\un x}_{0{\tilde 0}}, {\un z}, {\un b}_2, Y - y_2 )
 \right) \, {\hat L}_{k_2} ({\un z})
\een
\be\label{2Glin3}
\, \left( \int \, d^2 b_1 \, 
\frac{1}{\nabla^4_w}  n_1 ({\un z}, {\un w}, {\un b}_1, y_2 - y_1 )
\right) \, {\hat L}_{k_1} ({\un w}) \, n ({\un w}, {\un B}, y_1).
\ee
\eq{2Glin3} has the structure of three BFKL ladders (two factors of 
$n_1$ and one factor of $n$) with two Lipatov vertices, which are
responsible for production of gluons, inserted between them. It is
equivalent to the $k_T$-factorization prediction for two-gluon
production from a single BFKL ladder [see Eq. (30) in
\cite{braun}]. We have, therefore, proven that at the leading twist 
(large $k_T$) level, our two-gluon inclusive production cross section
(\ref{2Gincl}) reduces to the conventional $k_T$-factorized
expression (\ref{2Glin3}).

Two comments are in order here. First of all, it is a little worrisome
that in order to recover the conventional leading twist result of
\eq{2Glin3} we had to expand the initial conditions for the evolution of 
quadrupole amplitude $Q$ given by \eq{Q0} to the lowest order, as
shown in \eq{Q0n0}. Of course, by doing so, we have shown that the
leading twist formula (\ref{2Glin3}) is included in our full
expression (\ref{2Gincl}). Nevertheless, taking a solution of an
evolution equation at very short transverse distances, does not
necessarily imply doing the same to initial conditions of
evolution. For instance, if we are interested in the dipole amplitude
$N ({\un x}_0, {\un x}_{\tilde 0}, Y)$ at small ${\un x}_{0{\tilde
0}}$ we have to solve the linear part of \eq{eqN} with the full
initial conditions given by \eq{N0} and not with the leading order
initial conditions given by \eq{n0}. The kernel of \eq{eqN} involves
integration over all transverse sizes including large sizes where
multiple rescatterings are important and have to be included.
Multiple rescatterings become important at lower energies than the
small-$x$ evolution and thus have to be included as initial condition
even for linear (BFKL) evolution equation. (As one can show, multiple
rescatterings in the quasi-classical limit become important at
rapidity $y_{mult} \sim \ln 1/\as$, while the BFKL evolution becomes
important at $y_{BFKL} \sim 1/\as$.) Thus, expanding the initial
conditions of \eq{N0} would not be justified if one is interested in
the small ${\un x}_{0{\tilde 0}}$ of the amplitude $N ({\un x}_0, {\un
x}_{\tilde 0}, Y)$. The effects of saturation in the initial
conditions on the short distance/large $k_T$ behavior of the
amplitudes and gluon production cross sections have been studied
before in \cite{kkt1,klm}. It is exactly these effects that bring in
suppression of the nuclear modification factor $R^{pA}$ for the gluon
production \cite{klm,kkt1,aaksw}. Therefore, taking the large $k_1$ and
$k_2$ limit of \eq{2Gincl} more carefully may result in an expression
different from the $k_T$-factorization formula of
\eq{2Glin3}.

The second observation one has to make is that by now the reader can
appreciate the tremendous simplifications one needs to make
[e.g. \eq{Q0n0}, linearization of all evolution equations, etc.] in
order to recover the $k_T$-factorization formula of \eq{2Glin3}. This
is strikingly different from the case of single inclusive gluon
production considered in \cite{kt}. There the obtained expression for
the cross section was cast in a $k_T$-factorized form without making
any linearization assumptions, i.e., without taking the leading twist
(high-$k_T$) limit. This $k_T$-factorization result of \cite{kt} was,
indeed, unexpected and very puzzling. However it is also interesting
to observe that it does not hold for the double inclusive gluon
production cross section (\ref{2Gincl}). This leaves us guessing
whether the preservation of $k_T$-factorization in the formula from
\cite{kt} for the single gluon inclusive production cross section 
after multiple rescatterings and small-$x$ evolution had been included
(see \eq{1Gincl} above) was just incidental. A similar breakdown of
$k_T$-factorization has been observed recently for $q\bar q$
production in $pA$ collisions \cite{bgv}.

\section{Valence Quark-Gluon Production in Proton-Nucleus Collisions}

\begin{figure}
\begin{center}
\epsfxsize=5 cm
\leavevmode
\hbox{\epsffile{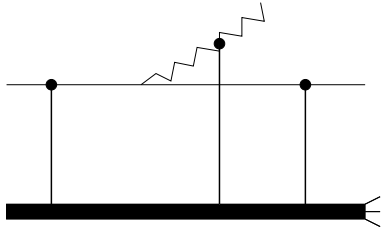}}
\end{center}
\caption{Production of a quark and a gluon including multiple scattering from 
the target.}
\label{fig:qA-qgx}
\end{figure}

\begin{figure}
\begin{center}
\epsfxsize=7 cm
\leavevmode
\hbox{\epsffile{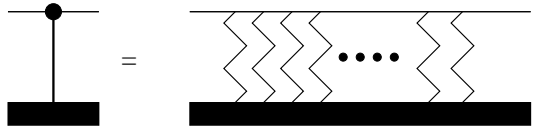}}
\end{center}
\caption{Multiple scattering of a quark or gluon on a target.}
\label{fig:mulscatt}
\end{figure}

In this section we calculate the cross section for production of a
valence quark and a gluon in high energy proton nucleus collisions.
Both produced quark and gluon assumed to have similar rapidity and to
be in the proton (deuteron) fragmentation region. In this case, one
can treat the proton (deuteron) as a dilute system of partons while
the target nucleus is treated as a Color Glass Condensate. The
produced quark and gluon then fragment into jets which can be
measured. Previously, this approach has been used to calculate valence
quark, photon and di-lepton cross sections in proton
(deuteron)-nucleus collisions \cite{adjjm,jjm1}. Here, we extend this
formalism to production both of a quark and a gluon. Explicitly, we
calculate the differential cross section for the following process:
\be
q(p)\,A \rightarrow q(q)\, g(k)\, X
\label{eqn:qA}
\ee
given by the amplitude 
\be
{\cal M} (q,k;p) \equiv <q(q)\, g(k)_{out} | q(p)_{in}> = 
<0_{out}| a_{out}(k) b_{out}(q) b^{\dagger}_{in}(p) | 0_{in}>
\label{eq:amp}
\ee
which, using the LSZ reduction formalism, can be written as (we set
the renormalization factors equal to $1$ since we are working at the
leading order in $\alpha_s$)
\be
{\cal M} & = &g \int d^4x \,d^4y \,d^4z \,d^4r \, 
d^4\bar{r} \,e^{i(q\cdot z + k\cdot r - p\cdot y)} 
\bar{u}(q) [i\stackrel{\rightarrow}{\slpartial}_z ] S_F (z,x) \gamma^{\nu} 
t^c S_F(x,y)  [i\stackrel{\leftarrow}{\slpartial}_y ] u(p) \nonumber \\
&&G_{\nu\rho}^{cb}(x,\bar{r}) D^{\rho\mu}_{ba}(\bar{r},r) \, \epsilon_{\mu}(k)
\label{eq:ampexp}
\ee
where $S_F$, $G_{\nu\rho}$ are the quark and gluon propagators in the
classical field background and $D^{\rho\mu}$ is defined such that
\be
\int d^4 r \,G^{0cb}_{\nu\rho} (x,r)\, D^{\rho\mu}_{ba} (r,y) \equiv 
\delta^c_a \, \delta_{\nu}^{\mu} \, \delta^4 (x-y)
\ee
where
$G_{\nu\rho}^0$ is the free gluon propagator.  This amplitude is shown
in Fig. \ref{fig:qA-qgx} where the quark and gluon lines with a thick
dot represent the propagators in the background field as illustrated
in Fig. \ref{fig:mulscatt}.
 
  \vspace{0.5in}

To proceed further, we write the the propagators in the above amplitude 
in momentum space. The amplitude is 
\be
{\cal M} = g \int {d^4k_1 \over (2\pi)^4} {d^4k_2\over (2\pi)^4} 
{d^4k_3\over (2\pi)^4} \bar{u}(q) \stackrel{\rightarrow}{\slq}
S_F(q,k_1) \,\gamma^\nu \, t^c \, S_F(k_2,p) \stackrel{\leftarrow}{\slp}
u(p) \, G_{\nu\rho}^{cb} (k_2-k_1,k_3) \,D^{\rho\mu}_{ba} (k_3,k)
\,\epsilon_{\mu} (k)
\ee

The quark and gluon propagators in the classical background
field are already known \cite{hw,ayala}. It is useful to separate the
free and interacting parts of the propagator in the following. Therefore,
we Define the interaction part of the propagators in momentum space as (and 
suppressing the color factor for the moment) 

\be 
S_F(q,p)&\equiv &(2\pi)^4 \delta^4(p-q)\, S^0_F
(p) + S_F^0 (q) \,\tau_f (q,p) \, S_F^0 (p) \nonumber \\ 
G^{\mu\nu}(q,p)&\equiv&
(2\pi)^4 \delta^4(p-q)\, G^{0\mu\nu} (p) + G^{0\mu}_{\rho}(q) \,
\tau_g (q,p)\, G^{0\rho\nu}(p)
\label{eq:decomp}
\ee
where the free propagators are 
\be
S_F^0 (p) &=& i{\slp \over p^2} \,\,\,\,\,\,\,\, and \,\,\,\,\,\,\,\,\,\,
G^0_{\mu\nu}(k) = {i\over k} \bigg[-g_{\mu\nu} + 
{\eta_\mu k_\nu + \eta_\nu k_\mu  \over \eta \cdot k}\bigg]
\ee
and $\eta_\mu$ is the light cone gauge vector so that $\eta \cdot A
\equiv A^- = 0$ defines the gauge we are working in. In this gauge the interaction 
part of the gluon propagator in \eq{eq:decomp}, denoted here by
$\tau_g (q,p)$, is diagonal in Lorentz indices, i.e., is proportional
to $g_{\mu\nu}$, which allowed us to suppress the Lorentz indices and
write it in the form shown in \eq{eq:decomp}. Such decomposition may
not hold in other gauges.  Inserting Eqs. (\ref{eq:decomp}) into the
amplitude and defining
\be
{\cal M}(q,\lambda ,k;p)\equiv M_1 + M_2 + M_3 + M_4 = 
\epsilon_\mu^{(\lambda)} (k) \,[M_1^\mu + M_2^\mu + M_3^\mu + M_4^\mu]
\label{eq:amp_dec}
\ee
where $\epsilon_\mu^{(\lambda)} (k) $ is the polarization vector of the 
produced gluon. we get 
\be
M_1 &=& -g \,\bar{u}(q)\, \slepsilon \,t^a\,S_F^0(q+k)\, 
\tau_f(q+k,p)\,u(p) \\
M_2 &=& -g\, \bar{u}(q)\, \tau_f(q,p-k)\, S_F^0(p-k)\, \slepsilon \, 
t^a\, u(p)\\
M_3 &=& -g \,\bar{u}(q) \, \gamma_{\nu} t^b\,\tau_g^{ba}(k,p-q)\,u(p)\, 
G_0^{\nu\mu}(p-q)\, \epsilon_\mu (k)\\
M_4 &=& -g \, \int d^4l \,\bar{u}(q) \,\tau_f(q,p-l)\,S_F^0(p-l)\,\gamma_{\nu}
\,t^b\,\tau_g^{ba}(k,l)\,u(p) \, G_0^{\nu\mu}(l)\, \epsilon_\mu (k) 
\label{eq:amp_1234}
\ee
where $\tau_f$ and $\tau_g$ are given by
\be
\tau_f (q,p)\equiv (2\pi)\delta(p^- - q^-) \,\gamma^-\, \int d^2x_t\, 
e^{i (q_t - p_t)\cdot x_t}\, [V(x_t) - 1]\\
\tau_g (q,p) \equiv 2p^- \, (2\pi)\delta(p^- - q^-) \, \int d^2x_t\, 
e^{i (q_t - p_t)\cdot x_t}\, [U(x_t) - 1].
\label{eq:UVdef}
\ee
The matrices $V$ and $U$ include all the multiple scatterings of the
quark and gluon as they propagate in the strong classical field of the
target and are given by
\be
V(x_t) &\equiv& \hat{P} \,e^{ig\int dz^-\, A^+_a (x_t,z^-) \,t_a} \label{vdef} \\
U (x_t) &\equiv& \hat{P} \, e^{ig\int dz^-\, A^+_a (x_t,z^-) \, T_a} \label{udef}.
\ee
Here $t_a$ and $T_a$ are matrices in the fundamental and adjoint
representations of the $SU(N)$ group respectively and $A^+_a (x_t,x^-)
= -g \,\delta (x^-) { \rho_a (x_t) \over \partial_t^2}$.

\begin{figure}
\begin{center}
\epsfxsize=10cm
\leavevmode
\hbox{\epsffile{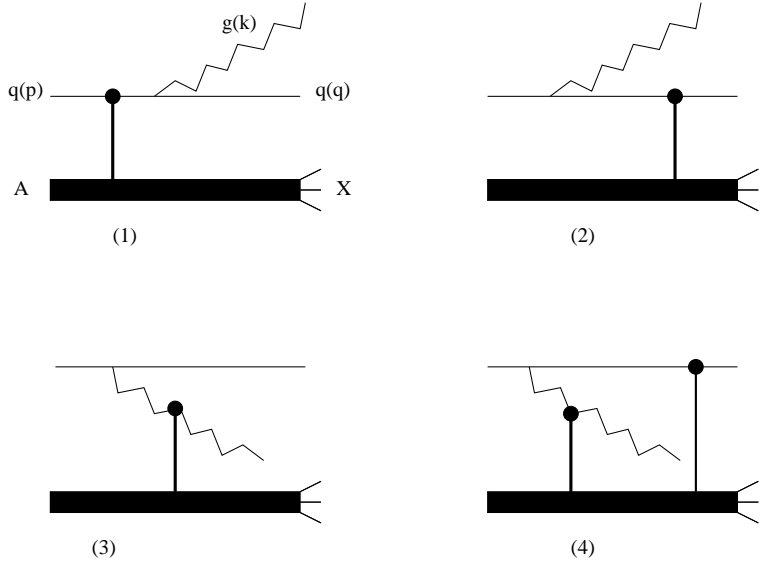}}
\end{center}
\caption{Diagrams corresponding to $M_1,M_2,M_3,M_4$.}
\label{fig:qgx_1234}
\end{figure}

With these definitions at hand, extracting an explicit factor of
$(2\pi) \delta(p^- - q^- -k^-)$ while using the delta function $\delta
(l^- - k^-)$ to do the $l^-$ integration and performing the $l^+$
integration in $M_4$ via contour integration using the $(p-l)$ pole,
we can write the amplitude as
\be
 M_1 &=&- i g {1 \over 2q\cdot k}  \bar{u}(q)\, \slepsilon\, 
(\slq + \slk)\,\gamma^-\,u(p)\,
 t^a\,[V(q_t + k_t) - (2\pi)^2 \delta^2(q_t + k_t)]\nonumber\\
M_2 &=& i g {1 \over 2p\cdot k}  \bar{u}(q)\, \gamma^- \, 
(\slp - \slk)\,\slepsilon \,u(p)\,
 [V(q_t + k_t) - (2\pi)^2 \delta^2(q_t + k_t)]\,t^a\nonumber\\
M_3 &=& i g {k^- \over p\cdot q}  \bar{u}(q)\, \gamma_{\nu}\,u(p)\, 
d^{\nu\mu} (p-q)\, \epsilon_\mu (k) \,t^b\,[
U^{ba}(q_t+k_t) - \delta^{ba}\,(2\pi)^2 \delta^2(q_t + k_t)]\nonumber\\ 
M_4 &=& i g {k^- \over p^-} \int  {d^2l_t \over (2\pi)^2} \bar{u}(q) 
\,\gamma^-\,(\slp - \sll)\,\gamma_{\nu}\,u(p)\,
{d^{\nu\mu} (l) \over l_t^2} \, \epsilon_\mu (k) \nonumber \\
&& [V(q_t + l_t) - (2\pi)^2 \delta^2(q_t + l_t)]
t^b [U^{ba}(k_t - l_t) - \delta^{ba}\,(2\pi)^2 \delta^2(k_t - l_t)]
\label{eq:amp_1234_mom}
\ee
where $d^{\mu\nu} (l) $ is related to the free gluon propagator via $
d^{\mu\nu} (l) \equiv - i\,l^2\, G_0^{\mu\nu}(l)$ and $l^- = k^-$, $l^+
= - {l_t^2 \over 2q^-}$. We have also set the transverse momentum of
the incoming quark to zero without any loss of generality. We show the
different diagrams contributing to the amplitude in
Fig. \ref{fig:qgx_1234}. They correspond, respectively, to the quark
multiply scattering from the target before or after radiating a gluon
in Figs. \ref{fig:qgx_1234}-1 and \ref{fig:qgx_1234}-2 and the
radiated gluon multiply scattering from the target in
Fig. \ref{fig:qgx_1234}-3 while in Fig. \ref{fig:qgx_1234}-4 both the
radiated gluon and the final state quark multiply scatter from the
target. For the sake of clarity, momenta of the incoming quark and
outgoing quark and gluon are shown explicitly in
Fig. \ref{fig:qgx_1234}-1.

\begin{figure}
\begin{center}
\epsfxsize=5cm
\leavevmode
\hbox{\epsffile{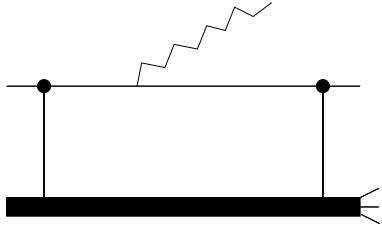}}
\end{center}
\caption{A typical diagram which is suppressed in the high energy limit 
and, therefore, not included.}
\label{fig:qgx_supp}
\end{figure}

In Fig. \ref{fig:qgx_supp} we show one of the diagrams which are
suppressed in the high energy limit and do not contribute and
therefore, not included in Eqs. (\ref{eq:amp_1234_mom}).  This is due
to the fact that the typical time scale for gluon emission is much
longer than the time between rescatterings in the target nucleus. The
diagram in \fig{fig:qgx_supp} is, therefore, suppressed by a power of
center of mass energy and can be safely neglected. Another diagram
(not shown), which is suppressed in the high energy limit for the same
reason, is when both the initial and final state quark lines as well
as the radiated gluon multiply scatter from the target.

To calculate the cross section, we need to square the amplitude
$|{\cal M}|^2$ (\ref{eq:amp_dec}). There is a factor of $-g_{\mu\nu} +
({1\over \eta \cdot k}) [k_\mu \eta_\nu + \eta_\mu k_\nu]$ coming from
squaring and summing over the polarization of the final state gluon in
the light cone gauge which can be used to simplify the
expressions. Furthermore, we define $z=q^-/p^-$ so that $1-z =
k^-/p^-$. Below, we list the different contributions coming from
squaring the amplitude. For reasons which will become clear shortly,
we consider the square of $M_1 + M_2$ first.  The contribution of the
$-g_{\mu\nu}$ term to the squared amplitude is (extracting a factor of
$g^2$ for convenience)
\be
-g_{\mu\nu} (M_1^\mu + M_2^\mu)^{\dagger} (M_1^\nu + M_2^\nu) 
&=& 16 p^-p^- \Bigg\{ 
{z(1-z)^2 \over [zk_t - (1-z)q_t]^2}\nonumber \\  
Tr [V^{\dagger}(q_t &+& k_t) - (2\pi)^2 \delta^2(q_t + k_t)]  t^a \, t^a 
[V(q_t + k_t) - (2\pi)^2 \delta^2(q_t + k_t)] \nonumber\\
+ {z(1 - z)^2 \over k_t^2} \
Tr [V(q_t &+& k_t) - (2\pi)^2 \delta^2(q_t + k_t)]  t^a \, t^a 
[V^{\dagger}(q_t + k_t) - (2\pi)^2 \delta^2(q_t + k_t)]  \nonumber\\
+  \Bigg[
(1-z)^2 (1 &+& z^2) {q_t^2 \over k_t^2  [zk_t - (1-z)q_t]^2} + {z^2(1-z^2) 
\over  [zk_t - (1-z)q_t]^2}
- {1-z^2 \over k_t^2}\Bigg] \nonumber\\ Tr \, t^a [V^{\dagger}(q_t &+&
k_t) - (2\pi)^2 \delta^2(q_t + k_t)] t^a [V(q_t + k_t) - (2\pi)^2
\delta^2(q_t + k_t)]\Bigg\}
\label{eq:m_12}
\ee
where $Tr$ denotes trace of color matrices. This term is identical, up
to color matrices, to the photon $+$ quark production calculated in
\cite{jjm2}. We now consider contribution of the 
$[k_\mu \eta_\nu + \eta_\mu k_\nu]$ piece. Using the Dirac equation and the
identity $\bar{u}(q) \gamma^- u(p) = 2 \sqrt{2p^-q^-}$  
(valid for on mass shell particles) simplifies the trace algebra 
considerably and we get 
\be
{[k_\mu \eta_\nu + \eta_\mu k_\nu]\over  \eta \cdot k} 
(M_1^\mu + M_2^\mu)^{\dagger} (M_1^\nu + M_2^\nu) &=&
32 \,p^- p^- z^2 \bigg[V^\dagger (q_t + k_t) t^a - 
t^a V^\dagger (q_t + k_t) \bigg]
\nonumber\\
&&\bigg[ {1 \over [zk_t -(1-z) q_t]^2} t^a V(q_t + k_t) - 
{1\over k_t^2} V(q_t + k_t) t^a \bigg].
\ee
Note that this piece has no analog in QED and would vanish in the case of 
photon emission.

A few remarks regarding the difference between photon and gluon
radiation is in order here. In single inclusive photon production in
$pA$ collisions as considered in \cite{jjm2}, the photon is emitted by
a quark scattering, via multiple gluon exchanges, from a target which
is treated as a classical gluon field, generated by recoilless sources
of color charge. The photon current is conserved and satisfies $k_\mu
\, M^\mu =0$ due to gauge invariance. It is therefore enough to work
in the covariant gauge where the sum over polarization of photons is
just $-g_{\mu\nu}$.  There is an essential difference between photon
and gluon radiation here due to the fact that in the case of gluon
radiation, one also needs to consider radiation of gluons from the
target and not just from the quark. This is essential for gauge
invariance of the amplitude and current conservation. However, it can
be shown that as long as one works in the light cone gauge, the gluon
radiation from the target vanishes identically. This is the case here
since we are working in the light cone gauge. However, this means that
one needs to keep the full projector $-g_{\mu\nu} + ({1\over \eta
\cdot k}) [k_\mu \eta_\nu + \eta_\mu k_\nu]$ rather than only the
$-g_{\mu\nu}$ piece which would be the case in the covariant gauge. We
now consider the rest of the diagrams:
\be
|M_3^{\dagger}M_1| &=& 16 \,p^-p^- \, z(1 + z^2) 
{q_t^2 -z q_t \cdot (q_t+k_t) \over q_t^2 [zk_t - (1-z)q_t]^2}
[U^{\dagger ab}(q_t + k_t) - \delta^{ab}(2\pi)^2 \delta^2(q_t + k_t)] 
\nonumber\\
&&Tr \, t^b\, t^a [V(q_t + k_t) - (2\pi)^2 \delta^2(q_t + k_t)] \nonumber\\
|M_3^{\dagger} M_2|&=& 16 \,p^-p^- z (1 + z^2) 
{q_t\cdot k_t \over q_t^2 k_t^2} 
 [U^{\dagger ab}(q_t + k_t) - \delta^{ab}(2\pi)^2 \delta^2(q_t + k_t)] 
 Tr\,t^b [V(q_t + k_t) - (2\pi)^2 \delta^2(q_t + k_t)]\,t^a \nonumber\\
|M_3|^2 &=& 16 p^- p^- {z(1+ z^2) \over q_t^2} 
[U^{\dagger ab}(q_t + k_t) - \delta^{ab}(2\pi)^2 \delta^2(q_t + k_t)] 
[U^{ca}(q_t + k_t) - \delta^{ca} (2\pi)^2 \delta^2(q_t + k_t)] \nonumber\\
&&Tr\,  t^b \, t^c \nonumber\\ 
|M^{\dagger}_3 M_4| &=&- 16\, p^-p^- z (1 + z^2) \int {d^2l_t \over (2\pi)^2} 
{q_t\cdot l_t  \over q_t^2 l_t^2} \,
[U^{\dagger ab}(q_t + k_t) - \delta^{ab}(2\pi)^2 \delta^2(q_t + k_t)]  
\nonumber\\
&&[U^{ca}(k_t - l_t) - \delta^{ca}(2\pi)^2 \delta^2(k_t - l_t)] \,
Tr\,t^b [V(q_t + l_t) - (2\pi)^2 \delta^2(q_t + l_t)] t^c 
\nonumber\\
|M_4^{\dagger} M_1| &=& - 16 p^-p^- z (1 + z^2) \int {d^2l_t \over (2\pi)^2}
{(1 - z) q_t \cdot l_t - z k_t\cdot l_t \over l_t^2 [zk_t - (1-z)q_t]^2} 
[U^{\dagger ab}(k_t - l_t) - \delta^{ab}(2\pi)^2  \delta^2(k_t - l_t)]
\nonumber\\  
&&Tr\, t^b [V^{\dagger }(q_t + l_t) - (2\pi)^2 \delta^2(q_t + l_t)]  t^a 
[V(q_t + k_t) - (2\pi)^2 \delta^2(q_t + k_t)] \nonumber\\
|M_4^{\dagger} M_2| &=& -16 p^-p^- z (1 + z^2) \int {d^2l_t \over (2\pi)^2} 
{k_t\cdot l_t \over l_t^2 k_t^2} 
[U^{\dagger ac}(k_t - l_t) - \delta^{ac}(2\pi)^2  \delta^2(k_t - l_t)] 
 \nonumber\\
&&Tr \,t^c  [V^{\dagger }(q_t + l_t) - (2\pi)^2 \delta^2(q_t + l_t)]  
[V(q_t + k_t) - (2\pi)^2 \delta^2(q_t + k_t)] t^a \nonumber \\
|M_4|^2 &=& 16 p^-p^- z (1 + z^2) 
 \int {d^2l_t \over (2\pi)^2} {d^2\bar{l}_t \over (2\pi)^2} 
{l_t\cdot \bar{l}_t \over l_t^2 \bar{l}_t^2} 
[U^{\dagger ac}(k_t - \bar{l}_t) - \delta^{ac}(2\pi)^2 
\delta^2(k_t - \bar{l}_t)] 
\nonumber\\
&&[U^{ab}(k_t - l_t) - \delta^{ab}(2\pi)^2  \delta^2(k_t - l_t)] 
 Tr  \, t^c t^b 
[V^{\dagger}(q_t + l_t) - (2\pi)^2 \delta^2(q_t + l_t)]\nonumber \\
&&[V(q_t + \bar{l}_t) - (2\pi)^2 \delta^2 (q_t + \bar{l}_t)].
\label{eq:m_34}
\ee
Note that for a given interference term such as $|M^{\dagger}_3 M_1|$,
there is also the conjugate term $|M^{\dagger}_1 M_3|$ which is obtained
from  $|M^{\dagger}_3 M_1|$ by daggering the color matrices. 
Eqs. (\ref{eq:m_12}-\ref{eq:m_34}) provide
the complete expression for the amplitude squared $|M|^2$. In order to
get the invariant cross section, one needs to include the phase space
and the flux factors given by ${d^3q \over (2\pi)^3} {1 \over 2q^-}$,
${d^3k \over (2\pi)^3} {1 \over 2k^-}$, and ${1 \over
2p^-}$. Including a factor of $1/2$ coming from averaging over the
incoming quark spin and restoring the coupling constant and the
overall delta function, the invariant cross section is given by
\be\label{QGincl}
q^-\,k^-{d\sigma^{qA\rightarrow qgX} \over d^3q\,d^3k} = {1\over 16 p^-} 
{1\over (2\pi)^6} 
(2\pi) \delta (p^- -q^- -k^-)\, g^2 \,|M|^2 .
\label{eq:cs_part}
\ee
This is the invariant cross section for production of a quark and
gluon in the scattering of a quark on a target nucleus (or a proton at
small $x$) including classical multiple scattering. In order to get
the invariant cross section for production of two hadrons or two jets
in a proton (deuteron)-nucleus collisions, one needs to convolute the
cross section given in Eq. (\ref{eq:cs_part}) with the (valence) quark
distribution function of a proton or deuteron and the quark or gluon
fragmentation functions
\be
E_{h1}E_{h2} {d\sigma^{pA\rightarrow h1\,h2\,X} \over d^3q_{h1}
d^3k_{h2}} \, = \, q_p (x_q) \otimes q^-\,k^- \,
{d\sigma^{qA\rightarrow qgX} \over d^3q\,d^3k} \otimes D^{q}_{h1}
(z_1) \otimes D^{g}_{h2} (z_2)
\label{eq:cs_2had} 
\ee
where $q_p(x_q)$ is the quark distribution function in a proton and
the quark and gluon fragmentation functions are denoted by $D^{q}_{h1}
(z_1)$ and $D^{g}_{h2} (z_2)$, while $\otimes$ denotes a convolution
over Bjorken $x$ for distribution function and over $z_1, z_2$ for
fragmentation functions. The cross section calculated here is valid
when one produces two hadrons (or jets) in the forward rapidity region
of a proton (deuteron) nucleus collision. It includes the effects of
quantum evolution (in $x$) in the target. To see this one has to
evaluate the Wilson lines ($U$'s and $V$'s) correlators in
Eqs. (\ref{eq:m_12}-\ref{eq:m_34}) using the
JIMWLK evolution equation \cite{jimwlk}. One can use this cross
section in order to investigate two particle correlations
(back-to-back jets) in the RHIC forward rapidity region which can be
measured, at RHIC for example, by the STAR detector at rapidity
$y=3.8$.

In order to consider the case when one of the hadrons in produced in
the mid rapidity region, one needs to allow the possibility that the
gluon is radiated not from the valence quark directly, but from
anywhere along the (in principle, non-linear ) gluon cascade between
the valence quark and the target. 

At this point, it is worthwhile to make a connection between the
notations used in different sections since they may seem disjoint to a
casual reader. The degrees of freedom are indeed the same even though
they are denoted differently due to convenience. In Sections II and
III the forward scattering amplitude of a quark--anti-quark dipole on
the target is denoted by $N (x_t,y_t)$ and can be expressed as
\be\label{nv}
N (x_t,y_t) \, = \, 1 - {1\over N_c} \ Tr \left< V^{\dagger} (x_t) V(y_t) \right>,
\ee
where $V$ is a path-ordered integral in the fundamental
representation, used in section $IV$ and defined in \eq{vdef}. Also,
the adjoint dipole amplitude denoted $N_G$ in, for instance, \eq{ngn},
is equal to
\be\label{ngu}
N_G (x_t,y_t) \, = \, 1 - {1\over N_c^2 -1} \ Tr \left< U^{\dagger} (x_t)
U(y_t) \right> ,
\ee
where $U$ is the path-ordered integral in the adjoint representation,
used in section $IV$ and defined in \eq{udef}. Furthermore, the
$S$-matrix of the color quadrupole interaction with the target, which
is denoted $Q_0 (x_t, y_t, z_t, r_t)$ in the classical case and
calculated in \eq{Q0} and denoted $Q (x_t, y_t, z_t, r_t)$ in the case
of quantum evolution included in \eq{Qev}, can be rewritten in terms
of the correlator of four path-ordered exponentials
\be\label{qv}
Q (x_t, y_t, z_t, r_t) \, = \, \frac{1}{N_c} \ Tr \left< V^{\dagger}
(x_t) V (z_t) V^{\dagger} (r_t) V (y_t) \right> .
\ee  
The relations in Eqs. (\ref{nv}), (\ref{ngu}) and (\ref{qv}) between
$N$, $N_G$, $Q$ and the correlators of $V$'s and $U$'s hold even when
the quantum evolution is included.

\section{Conclusions}

In this paper we have calculated two cross sections for inclusive
two-particle production relevant for the $dAu$ run at RHIC and for the
upcoming $pA$ run at LHC. The cross section for two-gluon production
at mid-rapidity for DIS is given by \eq{2Gincl}. The expression in
\eq{2Gincl} includes all multiple rescatterings of the produced gluons
on the target, along with the non-linear small-$x$ evolution effects
\cite{bk}. Even though, unlike the single gluon inclusive production 
cross section of \cite{kt} [see \eq{1Gincl} above], our two-gluon
cross section in \eq{2Gincl} can not be cast in $k_T$-factorized form,
it can be easily generalized to the case of proton--nucleus ($pA$)
scattering. Following \cite{kt} we note that the probability $n_1
({\un x}_{0{\tilde 0}}, {\un x}_{12}, {\un b}, y)$ of finding a dipole
$12$ in the original dipole $0{\tilde 0}$ can be related to the
unintegrated gluon distribution $\phi ({\un q}, y)$ as
\be\label{phi1}
\int d^2 b \, \frac{1}{\nabla_{x_{12}}^2} \, n_1 ({\un x}_{0{\tilde 0}}, 
{\un x}_{12}, {\un b}, y) \, = \, \frac{(2 \pi)^2}{\bas} \, \int \,
d^2 q \, e^{i {\un q} \cdot {\un x}_{12}} \, \phi ({\un q}, y),
\ee
where the coefficient on the right hand side of \eq{phi1} has been
fixed in order for \eq{1Gincl} to be reducible to conventional
$k_T$-factorization form of \cite{lr}. The information about the
original dipole $0{\tilde 0}$ is now contained in its unintegrated
gluon distribution $\phi ({\un q}, y)$. \eq{phi1} makes generalization
of \eq{1Gincl} from DIS to $pA$ rather straightforward: instead of the
unintegrated gluon distribution function $\phi ({\un q}, y)$ of the
incoming dipole, one has to use a BFKL evolved unintegrated gluon
distribution $\phi ({\un q}, y)$ of the proton in
\eq{phi1}, and, consequently, in \eq{1Gincl}.

To repeat the above procedure for \eq{2Gincl} we have to devise a
generalization procedure for the probability of finding two dipoles
$n_2$ as well. To do that, let us first clarify the physical meaning
of \eq{phi1}. The $1/\nabla_{x_{12}}^2$ term in
\eq{phi1} is due to $\nabla_{x_{12}}^2$, which is usual to the
definition of unintegrated gluon distribution in terms of the dipole
amplitude (see Eq. (2) in \cite{kkt1}), and $1/\nabla_{x_{12}}^4$,
which is proportional to gluons' propagators in a two-gluon exchange
amplitude. Thus, in \eq{phi1} the gluon distribution is obtained from
the dipole probability $n_1$ by connecting two $t$-channel exchange
gluons to the dipole $12$ in it. Now, generalization of \eq{phi1} to
$n_2$ becomes manifest: one has to connect two exchange gluons to each
of the two produced dipoles. The final expression reads
\be\label{phi2}
\int d^2 b_{1{\tilde 1}} \, d^2 b_{2{\tilde 2}} \, 
\frac{1}{\nabla_{x_{1{\tilde 1}}}^2 \, \nabla_{x_{2{\tilde 2}}}^2} \, 
n_2 ({\un x}_{0}, {\un x}_{\tilde 0}, Y; {\un x}_{1}, {\un x}_{\tilde
1}, y_1, {\un x}_{2}, {\un x}_{\tilde 2}, y_2) \, \nonumber \\ = \,
\frac{(2 \pi)^4}{\bas^2} \, \int \, d^2 q \, d^2 l \, e^{i {\un q} \cdot 
{\un x}_{1{\tilde 1}} + i {\un l} \cdot {\un x}_{2{\tilde 2}}} \,
\phi_2 ({\un q}, Y-y_1; {\un l}, Y-y_2),
\ee
where $b_{1{\tilde 1}} = ({\un x}_1 + {\un x}_{\tilde 1})/2$,
$b_{2{\tilde 2}} = ({\un x}_2 + {\un x}_{\tilde 2})/2$, and $\phi_2
({\un q}, Y-y_1; {\un l}, Y-y_2)$ is the two-gluon distribution
function in the incoming dipole $0{\tilde 0}$, with the two gluons
having transverse momenta ${\un q}$ and ${\un l}$ and rapidities
$Y-y_1$ and $Y-y_2$ with respect to the projectile onium.

Analyzing \eq{2Gincl} one can see that for scattering on a large
nucleus both $n_1$ and $n_2$ come into \eq{2Gincl} integrated over
impact parameter(s), as employed in Eqs. (\ref{phi1}) and
(\ref{phi2}). Therefore, using Eqs. (\ref{phi1}) and (\ref{phi2}) one
can rewrite \eq{2Gincl} in terms of single and double unintegrated
gluon distributions $\phi$ and $\phi_2$. Taking these distributions
for a proton (deuteron) instead of the quarkonium would accomplish
generalization of \eq{2Gincl} to the case of $p(d)A$ scattering.

Eqs. (\ref{eq:m_12}-\ref{eq:m_34}) along with \eq{QGincl} give us a
production cross section for a valence quark and a gluon in the
forward rapidity direction in $p(d)A$ scattering. If the correlators
of Wilson lines in Eqs. (\ref{eq:m_12}-\ref{eq:m_34}) are averaged in
the Gaussian approximation \cite{kjkmw}, the obtained cross section
(\ref{QGincl}) would reduce to the quasi-classical result containing
multiple rescatterings only. (For an explicit evaluation of color
averaging of the Wilson lines using a Gaussian weight, see
\cite{kw,bgv2}.) If the Wilson lines are averaged with the weight
function obtained from solving the JIMWLK evolution equation
\cite{jimwlk}, than \eq{QGincl} would include the complete effects of
small-$x$ evolution as well.

Before we conclude we would like to make a comment about the
applicability of Eqs. (\ref{eqN}), (\ref{eq:m_12}-\ref{eq:m_34}) for
RHIC kinematics. Indeed, in deriving these equations, we have assumed
for simplicity that the gluons are widely separated in rapidity, $y_2
\gg y_1$. On the other hand we know that particle production at
mid-rapidity at RHIC appears to be better described by the
quasi-classical physics leading to Cronin enhancement.  Therefore, if
one of the produced particles is at forward rapidity with the other
one being at mid-rapidity our formulas would apply, though one would
not need to include the small-$x$ evolution between the target nucleus
and the particle produced at mid-rapidity, since there the physics is
quasi-classical. However, the suppression in $R^{dAu}$, which is most
likely caused by small-$x$ evolution, sets in already at rapidity
$\eta = 1$ and continues all the way up to the highest achievable
rapidity at RHIC
\cite{brahms-1,brahms-2,phobos}. That means quantum evolution describes 
physics at $\eta \ge 1$.  Therefore, if both of the produced particles
are at rapidity $\eta \ge 1$, say if $y_1 = 1$ and $y_2 = 3$, we can
still have a large rapidity interval between them, $y_2 \gg y_1$, and
have quantum evolution between the target and the gluon at $y_1$
included in Eqs. (\ref{eqN}), (\ref{eq:m_12}-\ref{eq:m_34}). 
Indeed, the upcoming $pA$ run at the LHC would have
a much wider rapidity window, where our results would be even more
applicable.

\section*{Acknowledgments}

We are grateful to Francois Gelis, Dima Kharzeev, Genya Levin, 
Al Mueller and Raju Venugopalan for stimulating and informative 
discussions.

J.J-M. is supported in part by the U.S. Department of Energy under
Grant No. DE-FG02-00ER41132.  The work of Yu. K. was supported in part
by the U.S. Department of Energy under Grant No. DE-FG02-97ER41014.

\renewcommand{\theequation}{A\arabic{equation}}
  \setcounter{equation}{0} 
  \section*{Appendix A}

\begin{figure}
\begin{center}
\epsfxsize=16.5cm
\leavevmode
\hbox{\epsffile{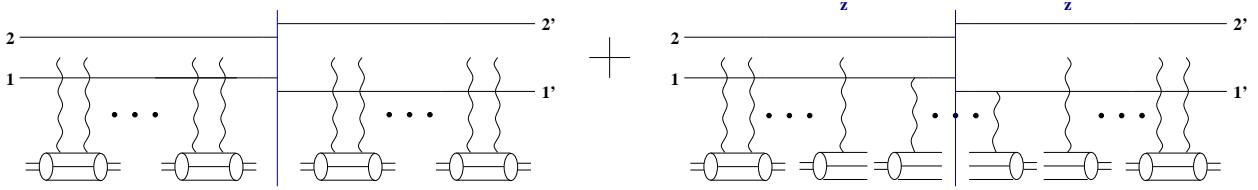}}
\end{center}
\caption{Leading diagrams contributing to the interaction of a color 
quadrupole $2,2',1,1'$ with a nuclear target in the large-$N_c$
limit.}
\label{quad}
\end{figure}

In this Appendix we calculate the $S$-matrix of the interaction a
quadrupole $2,2',1,1'$ with the target nucleus. The $S$-matrix
includes the Glauber-Mueller \cite{Mueller1,mv,km,KMc} multiple
rescatterings only and is denoted by $Q_0 ({\un x}_2, {\un x}_{2'},
{\un x}_1, {\un x}_{1'})$ in Sect. II above. The possible interactions
are shown in \fig{quad}. The first term there corresponds to the case
where all the interactions in the amplitude and in the complex
conjugate amplitude are virtual, i.e., each nucleon exchanges two
gluons with the $q\bar q$ pair and remains intact. This is the
diffractive piece of the interaction \cite{KMc}. The gluons connect to
both the quark and the antiquark lines. This is denoted by leaving the
gluons lines disconnected at the top ends. All the virtual exchanges
are leading at large $N_c$. The first diagram in \fig{quad} gives a
contribution
\be\label{1q}
e^{- x_{21}^2 \ln (1/x_{21} \Lambda) \, Q_{s0}^2 /4} \, 
e^{- x_{2'1'}^2 \ln (1/x_{2'1'} \Lambda) \, Q_{s0}^2 /4} 
\ee
which is just a product of the $S$-matrices of the dipoles $12$ and
$1'2'$.

The second diagram in \fig{quad} corresponds to the case of at least
one {\sl real} interaction: there the nucleon at longitudinal
coordinate $z$ interacts with the $q\bar q$ pair by a single gluon
exchange in the amplitude and in the complex conjugate amplitude. The
single gluon exchange breaks up the nucleon in the final state. We
will refer to this interaction as real \cite{km,yuri1}. The
interaction of the nucleon at $z$ is chosen to be the first real
interaction: all prior exchanges are virtual (exchanges to the left of
$z$ in the amplitude and to the right of $z$ in the complex conjugate
amplitude). After the interaction of nucleon at $z$ the exchanges can
be both real and virtual. However, in the large $N_c$ limit, only
those real exchanges contribute where gluons connect to either lines
$1$ and $1'$ or lines $2$ and $2'$. The color structure is similar to
the dipole model \cite{dip}: the nucleon at $z$ splits the original
single quark loop $2,2',1,1'$ into two, and the successive
interactions can only take place within each of the two resulting
loops.

The contribution of the second graph in \fig{quad} is therefore
\ben
\int_0^{L} \frac{dz}{L} \, e^{- \frac{1}{4} \, 
[ x_{21}^2 \ln (1/x_{21} \Lambda) 
+ x_{2'1'}^2 \ln (1/x_{2'1'} \Lambda) ] \, Q_{s0}^2 \, \frac{z}{L}} \, 
\left( - \frac{1}{4}  \, Q_{s0}^2  \right) \, 
[x_{22'}^2 \ln (1/x_{22'} \Lambda) 
+ x_{11'}^2 \ln (1/x_{11'} \Lambda) 
\een
\be\label{2q}
+ x_{21'}^2 \ln (1/x_{21'} \Lambda) + x_{2'1}^2 \ln (1/x_{2'1}
\Lambda)] \, e^{- \frac{1}{4} \, [ x_{22'}^2 \ln (1/x_{22'} \Lambda) +
x_{11'}^2 \ln (1/x_{11'} \Lambda) ] \, Q_{s0}^2 \, \frac{L - z}{L}}.
\ee
In \eq{2q} the first exponent resums all virtual interactions before
the first real interaction at $z$, the second exponent resums all the
real and virtual interactions with dipoles $22'$ and $11'$ following
the first real interaction, and the term in between accounts for the
first real interaction itself. We also average over the longitudinal
coordinate $z$ which varies from $0$ to $L$, where $L$ is the
longitudinal extent of the nucleus at a given impact parameter.

Performing the integration over $z$ in \eq{2q} and adding to it the
contribution from \eq{1q} yields \eq{Q0} in the text.

\end{document}